\documentclass[%
 aip,
 amsmath,amssymb,
 preprint,%
]{revtex4-1}

\usepackage{graphicx}
\usepackage{dcolumn}
\usepackage{bm}

\usepackage[utf8]{inputenc}
\usepackage[T1]{fontenc}
\usepackage{mathptmx}
\usepackage{etoolbox}

\usepackage{hyperref}
\hypersetup{
	colorlinks=true,
	linkcolor=blue,
	filecolor=blue,
	urlcolor=blue,
	citecolor=blue,
}
\usepackage{color}

  \usepackage{natbib}
\makeatletter
\def\@email#1#2{%
 \endgroup
 \patchcmd{\titleblock@produce}
  {\frontmatter@RRAPformat}
  {\frontmatter@RRAPformat{\produce@RRAP{*#1\href{mailto:#2}{#2}}}\frontmatter@RRAPformat}
  {}{}
}%
\makeatother
\begin{document}

\preprint{}

\title{New insights on carbon black suspension rheology  - anisotropic thixotropy and anti-thixotropy}
\author{Y. Wang}
\affiliation{ 
	Department of Mechanical Science and Engineering, University of Illinois at Urbana-Champaign, Urbana, Illinois 61801, USA}
\affiliation{ 
	Beckman Institute for Advanced Science and Technology, University of Illinois at Urbana-Champaign, Urbana, Illinois 61801, USA}
\affiliation{ 
	Joint Center for Energy Storage Research, Argonne National Laboratory, Lemont, Illinois 60439, USA
}
\author{R. H. Ewoldt}
  \altaffiliation[Author to whom correspondence should be addressed;]{ electronic email: ewoldt@illinois.edu}
\affiliation{ 
	Department of Mechanical Science and Engineering, University of Illinois at Urbana-Champaign, Urbana, Illinois 61801, USA}
\affiliation{ 
	Beckman Institute for Advanced Science and Technology, University of Illinois at Urbana-Champaign, Urbana, Illinois 61801, USA}
\affiliation{ 
	Joint Center for Energy Storage Research, Argonne National Laboratory, Lemont, Illinois 60439, USA
}
\affiliation{ 
	Materials Research Laboratory, University of Illinois at Urbana-Champaign, Urbana, Illinois 61801, USA
}

\date{\today}

\begin{abstract}
	 Abstract: We report a detailed experimental study of peculiar thixotropic dynamics of carbon black (CB, Vulcan XC-72) suspensions in mineral oil, specifically the observation of sequential stress increase then decrease at a fixed shear rate in a step-down test. We verify that such dynamics, though peculiar, come from a true material response rather than experimental artifacts. We also reveal how this long-time stress decay is associated with anti-thixotropy, rather than viscoelasticity, by using orthogonal superposition (OSP) rheometry to probe viscoelastic moduli during the step-down tests. The orthogonal storage and loss modulus are present, showing this two-timescale recovery then decay response, which demonstrates that this response is anti-thixotropic, and it involves shear-induced structuring. We further show a mechanical anisotropy in the CB suspension under shear using OSP. Based on the rheological results, a microstructural schematic is proposed, considering qualitatively thixotropic structure build-up, anti-thixotropic densification, and anisotropic structure evolution. Our observation for these CB suspensions is outside the standard paradigm of thixotropic structure-parameter models, and the elastic response provides us with new insight into the transient dynamics of CB suspensions.
\end{abstract}

\maketitle

\section{\label{sec:introduction}INTRODUCTION}

Understanding and predicting the rheological properties of carbon black (CB) suspensions are desired in many applications such as inks \cite{Khandavalli2018}, paints \cite{Hellinckx1969}, and semi-solid flow batteries \cite{Duduta2011,Fan2014,Wei2015,Parant2017,Cerbelaud2014,Mayer2020}. However, this is a challenging task because the structure of CB evolves in a complicated manner under flow. The fundamental building blocks of CB, the primary particles, are strongly fused by covalent bonds into fractal aggregates, and the individual aggregates join together by van der Waals forces to form agglomerates \cite{Usersbook}. At high shear rates, large CB agglomerates break into smaller sizes when the applied shear force is adequate to overcome the attractive force between them \cite{Usersbook}. This results in a lower hydrodynamic volume, and therefore, a lower viscosity \cite{Youssry2013}. At low shear rates, large but loosely-connected agglomerates are formed; for suspensions with volume fraction higher than the percolation threshold, this leads to a space-filling network \cite{Trappe2000,Trappe2001,DaSilvaLeiteCoelho2017,Richards2017} which greatly increases the flow resistance. Under certain shear conditions, these large, open, and loosely-connected agglomerates can self-organize and inter-penetrate \cite{Hipp2019}, due to the fractal nature of CB \cite{Narayanan2017}. Such densification decreases the viscosity of suspensions and has been reported in various flocculated systems \cite{Vermant2005, Hoekstra2003, Varadan2001}. At rest, the elastic modulus of CB suspensions can stiffen with the applied stress prior to yielding \cite{Osuji2008a}, due to the local rearrangement of CB aggregates \cite{Ngouamba2020}. Under shear, CB suspensions can also form anisotropic structures. For example, shear-induced log-rolling structures aligned in the vorticity direction at low dimensionless shear rates have been observed with CB \cite{Osuji2008,Negi2009,Grenard2011,Varga2019}. The structure of CB also shows shear history dependence, therefore, the suspensions can have a tunable yield stress \cite{Ovarlez2013}, elastic modulus \cite{Osuji2008a}, conductivity \cite{Helal2016}, and impedance \cite{Jiang2018} under different shear histories.

Many attempts have been made to characterize the structure of CB suspensions under flow for a better understanding of its rheology. Mewis and Schoukens \cite{Mewis1977} compared the mechanical spectra of CB suspensions under different shear rates and found that larger shear rate results in a smaller structural size in the CB network. Other characterization techniques that have been used include orthogonal superposition \cite{Sung2018}, ultrasonic \cite{Perge2014}, dielectric \cite{Richards2017}, and electric measurement \cite{Helal2016} coupled with rheology. Some direct measurements of the CB structure have been done using neutron scattering \cite{Hipp2021}, light scattering \cite{Bezot1999}, and optical microscopy \cite{Yearsley2012}, but most of these did not measure the transient response of CB suspensions under flow. Understanding the structure evolution of CB suspensions under flow through its transient response remains a complicated and unresolved problem, which is directly relevant to many flow-related applications.

\begin{figure*}
	\includegraphics[scale=1]{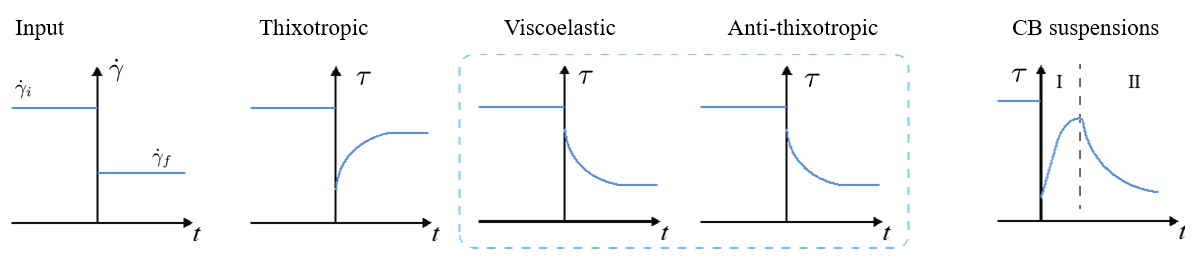}	
	\caption{Step shear rate tests can distinguish between thixotropy and anti-thixotropy, but not anti-thixotropy and viscoelasticity. For the step down input, $\dot \gamma$, possible evolutions of shear stress, $\tau(t)$, are shown: ideal thixotropy, viscoelasticity, and anti-thixotropy. More realistic stress responses of carbon black suspensions are shown as "CB suspensions", where some CB suspensions in literature show a stress growth (regime I, as observed in Ref.~\cite{Dullaert2005}), while others show a stress decay (regime II, as observed in Ref. \cite{Narayanan2017, Hipp2019}).}
	\label{Fig.thixo_anti-thixo_VE}
\end{figure*}

Furthermore, our understanding of the transient mechanical responses and structure changes under shear is greatly inhibited due to a lack of clear differentiation between thixotropy, anti-thixotropy, and viscoelasticity, while all of them have been reported on CB suspensions before \cite{Narayanan2017,Dullaert2006,Osuji2008}. Thixotropy is defined as ``the continuous decrease of viscosity with time when flow is applied to a sample that has been previously at rest and the subsequent recovery of viscosity in time when the flow is discontinued''  \cite{Dullaert2005,Dullaert2005b,Mewis2009,Barnes1997}. While this definition of thixotropy refers to a purely viscous phenomenon \cite{Larson2015,Rubio-Hernandez2020}, time-dependent structure breakdown or buildup generally impacts viscous, elastic, and plastic stresses. This may be called thixotropy, or structuration \cite{Grenard2014}, or aging and rejuvenation \cite{Larson2015, Sen2021, Mewis2009}. Thixotropy and viscoelasticity are usually differentiated using step-up in shear rate (also called break-down flow) and step-down in shear rate (also called build-up flow) tests \cite{Dullaert2005a,Dullaert2006}. Fig.~\ref{Fig.thixo_anti-thixo_VE} shows the possible shear stress evolutions of thixotropic, anti-thixotropic, and viscoelastic materials, in response to a step-down in shear rate. An ideal thixotropic fluid \cite{Mewis2009} would show an initial sudden decrease in stress, followed by a gradual increase until steady state is reached, when the applied shear rate is decreased from $\dot \gamma_i$ to $\dot \gamma_f$, because of structure build-up. For a viscoelastic fluid, on the other hand, the stress response is a monotonic decrease towards steady state \cite{Mewis2009}, because of elastic stress relaxation. The transient dynamics of CB suspensions are dominated by thixotropy, because the CB structure build-up and break-down processes mentioned earlier happen over timescales typically longer than the viscoelastic relaxation time \cite{Mewis2009, Dullaert2005}. The thixotropy of CB suspensions is well-studied using step-up and step-down in shear rate tests \cite{Dullaert2005,Dullaert2006}, creep \cite{Coussot2002,Grenard2014}, and flow sweeps \cite{Divoux2013}. For example, a suspension of one particular carbon black (FW2 from Degussa) in naftenic oil (Shell S6141) \cite{Dullaert2005} was observed to show thixotropic stress recovery with a step-down in shear rate test, as shown in the last column, "CB suspensions", in Fig.~\ref{Fig.thixo_anti-thixo_VE}, where the stress increases with time in regime I. 

Prior works have also claimed CB suspensions to show anti-thixotropy \cite{Ovarlez2013} (also called rheopexy, or negative thixotropy), as the shear stress decreases with time in a step-down in shear rate test \cite{Hipp2019} and increases with time in a step-up in shear rate test \cite{Narayanan2017}, as shown in Fig.~\ref{Fig.thixo_anti-thixo_VE}, the transient stress response of "CB suspensions" in regime II. The thixotropy and anti-thixotropy of CB suspensions are therefore contradictory results for the transient stress dynamics under step shear rate flow. However, the claims of CB being anti-thixotropic are based on observations that can be due to either anti-thixotropy or viscoelasticity. As shown in Fig.~\ref{Fig.thixo_anti-thixo_VE}, both anti-thixotropic and viscoelastic materials would show a monotonic decrease in shear stress to the steady state value. Therefore, a step in shear rate test cannot differentiate anti-thixotropy and viscoelasticity, although it is commonly used to distinguish between thixotropy and viscoelasticity. Fig.~\ref{Fig.thixo_anti-thixo_VE} only shows the case of step-down in shear rate, and the same is true for the step-up test. This results in an indistinguishable stress response for thixotropic, anti-thixotropic and viscoelastic materials \cite{Agarwal2021}. 

Orthogonal superposition (OSP) rheometry \cite{Vermant1998} can potentially help to distinguish between viscoelasticity and anti-thixotropy. This is performed by applying a step-down in shear rate flow in the primary (rotational) direction, and simultaneously superposing a small amplitude oscillatory flow in the orthogonal direction. Therefore, the evolution of orthogonal storage modulus, $G_\perp^\prime$, and orthogonal loss modulus, $G_\perp^{\prime \prime}$, under the step change in the main flow direction (the rotational flow) can be observed. The change in $G_\perp^\prime$ and $G_\perp^{\prime \prime}$ will give insight to distinguish viscoelastic and anti-thixotropic materials. This is because viscoelasticity involves storage and relaxation of elastic energy and stress, while anti-thixotropy is a result of change in structure, such as change in the size of agglomerates, which itself does not store energy, but affects the viscosity and moduli of suspensions. Due to the fundamental difference, we conjecture that the moduli under the step-down flow evolve differently for viscoelasticity and anti-thixotropy: for anti-thixotropic materials, both $G_\perp^\prime$ and $G_\perp^{\prime \prime}$ decrease with time, while for viscoelastic materials, the change of $G_\perp^\prime$ and $G_\perp^{\prime \prime}$ is material and model specific. Therefore, observing $G_\perp^\prime(t)$ and $G_\perp^{\prime \prime}(t)$ under step in shear rate flow provides us with new insight to distinguish between anti-thixotropy and viscoelasticity, and it can help us resolve the uncertainty in the origin of stress decay in CB suspensions. 

In this paper, we aim to experimentally study the transient dynamics of one particular CB, Vulcan XC-72, in heavy mineral oil suspensions under the step-down in shear rate flow, and resolve the ambiguity between thixotropy, anti-thixotropy, and viscoelasticity of CB suspensions. We do this by recording the shear stress, $\tau$, evolving with time during step-down in shear rate tests, when the applied shear rate is decreased from initial shear rate, $\dot \gamma_i$, to various final shear rates, $\dot \gamma_f$. A short-time recovery, followed by a long-time decay in $\tau$ is observed. Such peculiar decay in stress has been reported before \cite{Hipp2019}, but several hypotheses remain to be tested. Experimental artifacts must be checked, such as wall-slip, which can potentially lead to apparent stress decrease. The repeatability and reversibility of such peculiar stress decrease are tested, and we verify that such recovery-then-decay dynamics is a reversible, reproducible, and intrinsic material response. Moreover, whether the decrease in stress comes from anti-thixotropy or viscoelasticity remains unknown. With the help of OSP, we record the orthogonal moduli, $G_\perp^\prime$ and $G_\perp^{\prime \prime}$, under step-down tests. This helps us attribute the long-time decay to shear-induced anti-thixotropic structure rearrangement, rather than viscoelastic stress relaxation. The anti-thixotropy is also proved by hysteresis flow sweep. By comparing the results for shear stress and orthogonal moduli, we show a mechanical anisotropy in the thixotropic timescales and the magnitude of moduli of CB suspensions under shear. We further propose a micro-structural schematic, considering qualitatively thixotropic structure build-up, anti-thixotropic structure rearrangement, and anisotropic structure evolution. This paper is organized by first showing transient shear stress results in Sec.~\ref{sec:stepshear_results} and transient orthogonal moduli in Sec.~\ref{sec:osp_results}, followed by the anisotropy of CB suspensions, by comparing the responses in both directions, in Sec.~\ref{sec:timescale}. The concentration dependence and additional evidence from hysteresis are shown in Sec.~\ref{sec:concentration} and Sec.~\ref{sec:hysteresis} respectively. Discussion on the growth of anisotropy and mechanical schematics are shown in Sec.~\ref{sec:structure} and \ref{sec:anisotropy}.

\section{\label{sec:materials_methods}MATERIALS AND METHODS}

\subsection{\label{sec:shear}Shear rheological measurements}

Rheological measurements were performed at $\rm25^oC$ on a rate-controlled, separated motor-transducer rheometer (ARES-G2 from TA Instruments). A cone-and-plate geometry with a diameter of 25 mm and a cone angle of 0.1 rad was used, unless specified otherwise. There was no edge failure or sample degradation during the tests, verified by repeat measurements and videos of the sample edge during the tests, and there was no wall slip or shear banding during the tests, verified by a 25 mm parallel plate geometry at different measuring gaps \cite{Ewoldt2015}.

The behavior of CB suspensions is shear history dependent. Therefore, a 10 min preshear at 300 $\rm s^{-1}$ was applied before every measurement \cite{Grenard2014}, until the shear stress reached the steady state (within 3$\%$ variation). The preshear protocol is intended to erase the memory of any mechanical/shear history of the sample \cite{Jamali2022} and to ensure consistency across all measurements. CB samples were subjected to several rheological test procedures, each proceeded with the standard preshear protocol. These were: (i) step-down in shear rate tests, also called the structure recovery tests, to record the transient shear stress response; (ii) repeated step in shear rate tests using different geometries and gaps to verify the results; and (iii) hysteresis tests that provide additional evidence for anti-thixotropy. The details and results of each of these tests are described and discussed in Sec.~\ref{sec:stepshear_results} and Sec.~\ref{sec:hysteresis}.

\subsection{\label{sec:osp}Orthogonal superposition (OSP)}
Orthogonal superposition (OSP), in which a small amplitude oscillation is superposed in the direction perpendicular to that of the primary shear flow, is valuable in measuring shear-controlled structures \cite{Potanin1997, Mewis1977, Jacob2015, Colombo2017}, probing anisotropic structures \cite{Kim2013, Colombo2017}, and studying nonlinear viscoelasticity using linear concepts \cite{Vermant1998, Mewis2001, Kim2013}. While scattering, microscopy, and other methods mentioned earlier can characterize the structure of CB under flow, OSP is useful for studying the transient behavior and investigating the anisotropy in the mechanical properties of materials.

In this study, the transient orthogonal storage and loss moduli, $G_{\perp}^{\prime} (t; \omega, \gamma_0)$ and $G_{\perp}^{\prime \prime} (t; \omega, \gamma_0)$, were monitored as functions of time, $t$, by superposing a small amplitude oscillatory shear (SAOS) at a constant frequency, $\omega$, and amplitude, $\gamma_0$, in the orthogonal direction during a step-down in shear rate in the rotational direction. To quantify the mechanical anisotropy and its growth with shearing time in CB suspensions, a 2D-small amplitude oscillatory shear (2D-SAOS) \cite{Colombo2017} was applied, where an axial deformation $\gamma_\perp=\gamma_0 {\rm sin} (\omega t)$ was superposed on the conventional angular deformation $\gamma=\gamma_0 {\rm sin} (\omega t)$, and the peak amplitude is $\sqrt{2} \gamma_0$. A double gap slotted concentric cylinder OSP geometry on the ARES-G2 was used to apply simultaneous flows in the primary (rotational) and orthogonal (axial) directions. The diameters of the internal surface of the cup, internal surface of the bob, external surface of the bob, and external surface of the cup are 27.735, 29.403, 32.072, and 33.996~mm respectively; the geometry can be operated at immersed height between 42.655 to 38.655~mm. The consistency in measurements with different immersed heights are tested using Carbopol, as shown in the supporting information (Fig.~S1). For CB suspensions, an immersed height of 42.655 mm was used for all measurements in this study. Similar to shear rheology tests, a 10 min preshear at 300~$\rm s^{-1}$ in the rotational direction was applied before every measurement, until the shear stress reached the steady state (within 3$\%$ variation). The details and results of OSP tests are shown in Sec.~\ref{sec:osp_results} and Sec.~\ref{sec:anisotropy}.

\subsection{\label{sec:materials}Materials}
The carbon black used in this study is Vulcan XC-72 (Cabot Corporation), which has a density of 1,800~$\rm kg/m^3$. The Vulcan XC-72 CB is fractal, exhibiting a hierarchy of morphological features, and is well studied. The diameter of the primary particles is 20~nm, which are strongly fused by covalent bonds into aggregates with a fractal dimension $D_f = 2.7$ \cite{Richards2017}. The CB particles are dispersed in a heavy mineral oil (Fisher Scientific), which has a density of 1,200~$\rm kg/m^3$ and a Newtonian viscosity of  0.14~$\rm Pa \cdot s$ at 25~$\rm ^oC$, shown in Fig.~S2. The suspensions were prepared using a rotor-stator homogenizer (IKA) by shearing at 4000 rpm for more than 10 min to make a well dispersed and stable CB suspension in mineral oil. 

\begin{figure*}[ht]
	\begin{minipage}[]{0.45\textwidth}
		\centering
		\textbf{(a)}\par
		\includegraphics[scale=0.35,trim={0 0 0 0},clip]{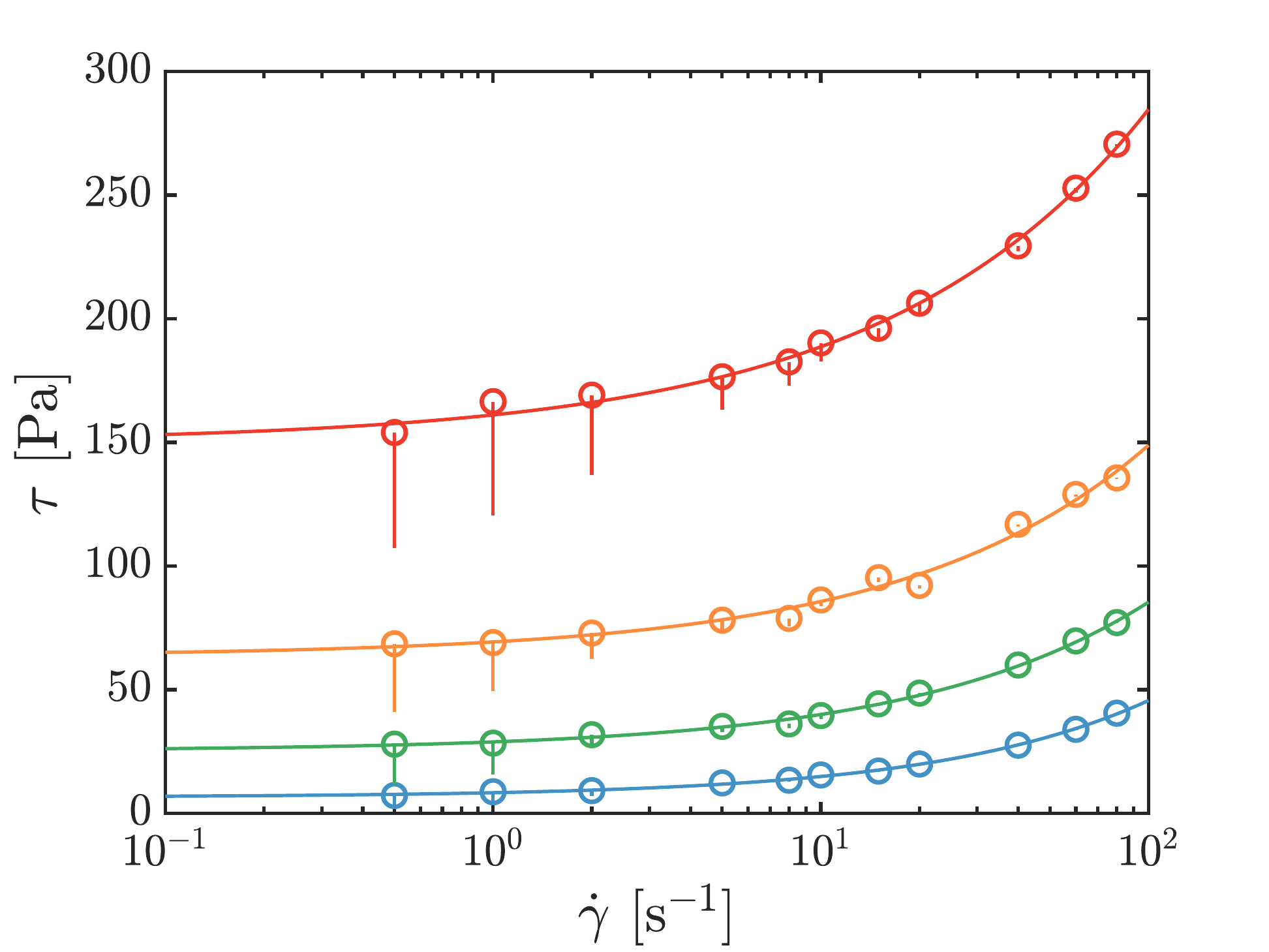}
	\end{minipage}
	\begin{minipage}[]{0.45\textwidth}
		\centering
		\textbf{(b)}\par
		\includegraphics[scale=0.48,trim={0 0 0 0},clip]{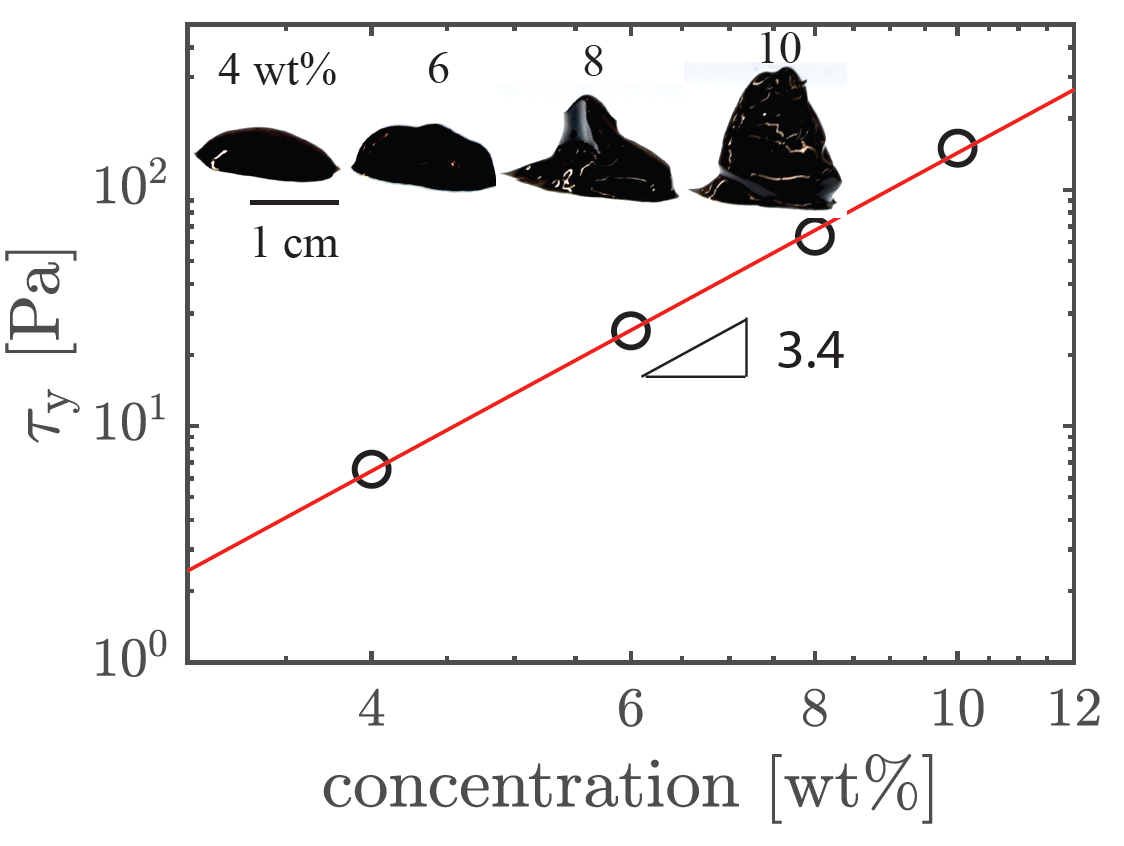}
	\end{minipage}
	\caption{Carbon black (CB) suspensions in heavy mineral oil with four different mass fractions: (a) flow sweep of CB suspensions with four mass fractions, from the top to bottom are 10, 8, 6, 4~wt$\%$ respectively. Symbols indicate the maximum transient stress during the step down shear tests, when the shear rate decreases from 100~$\rm s^{-1}$ to the shear rates of interest; the stress keeps decreasing after reaching the maximum value, as shown by the line segment below each symbol; full transient data shown in Fig. \ref{Fig.VulcanCB_6pctwt_stepshear_stress}. The solid lines show the Herschel-Bulkley fits to the open symbols. (b) The apparent yield stress, $\tau_y$, obtained from the Herschel-Bulkley model fits, versus the mass fraction; the red line is the power-law fit, which gives a slope of 3.4; the inset has photos of CB suspensions at four different mass fractions, with visual evidence of yield stress behavior.}
	\label{Fig.VulcanCB}
\end{figure*} 

In this study, four mass fractions of CB suspensions were prepared and tested, which are  4, 6, 8, 10~wt$\%$; the flow curves are shown in Fig.~\ref{Fig.VulcanCB}(a). With the increase in CB concentration, the yield stress increases, as suggested by the photos in Fig.~\ref{Fig.VulcanCB}(b) (inset). The flow curve in Fig.~\ref{Fig.VulcanCB}(a) was obtained by shearing the suspension at 100~$\rm s^{-1}$ until the steady state was reached, then stepping down to the shear rate of interest. This is to erase any mechanical history and to ensure that the suspension was at the same structural state at each shear rate of interest \cite{Hipp2019}, instead of sweeping up or down shear rates and collecting the steady value at each rate, as is conventionally done. At higher shear rates, stresses increased with time and reached steady state very quickly, and could be recorded within 10~s, which is in agreement with the thixotropic dynamics reported before \cite{Hipp2019}. At lower shear rates, however, no steady state was observed, and the stress started decreasing once it reached its maximum value and it kept decreasing even after very long times ($\sim$~3000~s). The absence of steady state has also been reported by Hipp $et~ al.$ \cite{Hipp2019}. In this case, the maximum stress observed during the transient increase-and-decrease was recorded and plotted as an open circle in the flow curve, as shown in Fig.~\ref{Fig.VulcanCB}(a). The vertical solid line extending below each circle shows the extent of subsequent stress decay in the time that follows. The full transient stress responses are shown in Fig.~\ref{Fig.VulcanCB_6pctwt_stepshear_stress}, which will be discussed in details later. From Fig.~\ref{Fig.VulcanCB}(a) we can see that at a given shear rate, the higher the concentration, the larger the decrease in stress. For the 4~wt$\%$ CB suspension (shown in blue), the decrease in stress is unnoticeable at shear rate of 10~$\rm s^{-1}$, while for the CB suspension of 10~wt$\%$ (shown in red), the stress decreases by around 40 Pa in 600 s at the same shear rate. For a given concentration, the higher the shear rate, the smaller the decrease in the stress. For the 10~wt$\%$ suspension, the decrease becomes unnoticeable when the shear rate is larger than around 80~$\rm s^{-1}$. Here, we use the maximum stresses during step-down in shear rate tests to fit the Herschel-Bulkley model \cite{Macosko1994, Nelson2017} for each concentration, 
\begin{equation}
\label{Eq.HB}
\tau = \tau_y \left[ 1+ \left(  \frac{\dot \gamma}{\dot \gamma_{\rm crit}} \right)^n \right],    ~~~~~~\rm for ~|\tau| > \tau_y,
\end{equation}
with $\tau$ being the stress and $\dot \gamma$ being the shear rate; $\tau_y$, $\dot \gamma_{\rm crit}$, and $n$ are fit parameters. The fit results for each concentration are shown in Tab.~\ref{tab:ss}, and the fitted apparent yield stress, $\tau_y$, is plotted in Fig.~\ref{Fig.VulcanCB}(b) as a function of concentration. Consistent with the photos (inset), the yield stress increases with concentration, and the increase follows a power law of slope 3.4, shown by the red solid line in Fig.~\ref{Fig.VulcanCB}(b). The value of the slope is consistent with what was found in previous literature for fractal CB suspensions, which reported a slope of 3.5 \cite{Grenard2014}.

\begin{table}
\caption{\label{tab:ss}Herschel-Bulkley model (Eq. \ref{Eq.HB}) fit results for CB suspensions with different concentrations.}
\begin{ruledtabular}
\begin{tabular}{cccc}
Concentration & $\tau_y$ & $\dot \gamma_{\rm crit}$ & $n$ \\
 wt$\%$ & Pa & $\rm s^{-1}$ &  \\
\hline
4 & 6.56 & 6.99 & 0.67 \\
6 & 25.30 & 24.33 & 0.61 \\
8 & 63.61 & 60.76 & 0.59 \\
10 & 150.00 & 122.06 & 0.54 \\
\end{tabular}
\end{ruledtabular}
\end{table}

\section{\label{sec:exp_results}Experimental RESULTS}

\subsection{\label{sec:stepshear_results}Verification of a reversible, reproducible, and intrinsic material response}

\begin{figure}[h!]
	\includegraphics[scale=.3]{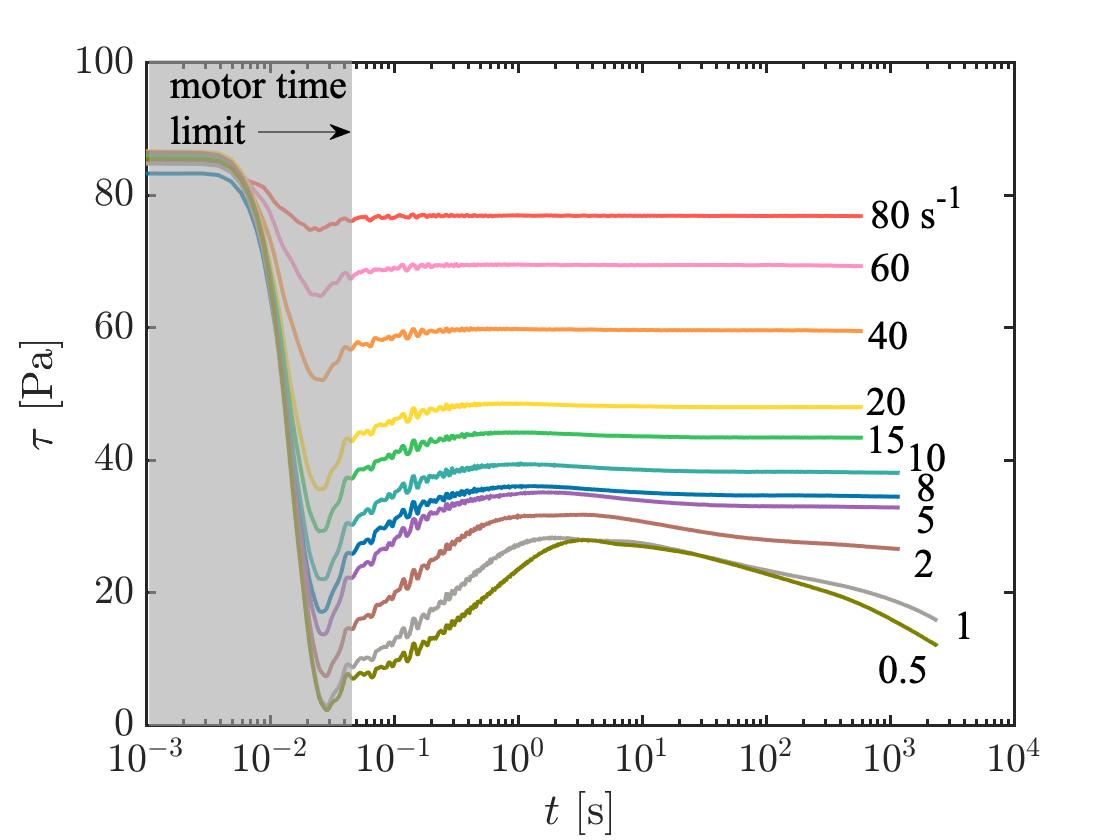}	
	\caption{The transient stress measured during the step-down shear tests with the initial shear rate of 100~$\rm s^{-1}$ and final shear rates ranging from 80 to 0.5~$\rm s^{-1}$ for the 6~wt$\%$ CB suspension shows a two-region response. At high shear rates, stress shows a quick thixotropic recovery, while at low shear rates, a peculiar, combined short-time recovery and long-time decay dynamics is observed.}
	\label{Fig.VulcanCB_6pctwt_stepshear_stress}
\end{figure}

A peculiar combined short-time recovery and long-time decay is observed in transient stress responses at certain constant applied final shear rates, $\dot \gamma_f$, on 6~wt$\%$ CB suspensions, as shown in Fig.~\ref{Fig.VulcanCB_6pctwt_stepshear_stress}. After preshearing at 300~$\rm s^{-1}$ for 10~min, the sample was sheared at an initial shear rate, $\dot \gamma_i=100 ~\rm s^{-1}$ for 2~min, and then the applied shear rate was stepped down from the initial value to different final shear rates ranging from 0.5 to 80~$\rm s^{-1}$, at time zero. Fig.~\ref{Fig.VulcanCB_6pctwt_stepshear_stress} shows that two distinct ranges are observed in transient stress responses. In the high final shear rate range ($\dot \gamma_f > 15 \rm ~s^{-1}$), the stress shows a typical fast thixotropic recovery: increasing with time and reaching steady state within 1~s. For the final shear rate of 80~$\rm s^{-1}$, the recovery is so fast (within motor time limits) and the stress recovered is so small that the thixotropy is negligible. In the low final shear rate range ($\dot \gamma_f < 15 \rm ~s^{-1}$), the stress shows a peculiar recovery-and-decay response as discussed before: increasing until reaching its maximum during the thixotropic recovery, then starting to decrease and never reaching steady state even after 2400~s. In an extreme case where $\dot \gamma_f = 1 \rm ~s^{-1}$, the steady state was not observed even after 10,000 s. The grey area at very short time (50~ms) represents the motor ramping-down time, before which the applied shear rate is unsteady because of a finite motor response time. The input shear rate time histories during the step-down tests are shown in Fig S3, where we can see that the motor ramping time varies for different final shear rates; the lower the final shear rate, the longer it takes for the motor to reach the steady velocity. The maximum time for the motor to reach the steady velocity is around 50~ms, guaranteeing that the transient stress responses outside the grey area are not due to motor transients.

The transient recovery-then-decay responses at constant applied shear rates also appear in CB suspensions at other concentrations (4, 8, 10~wt$\%$), and the detailed results are shown in Fig.~S4. The critical shear rate, at which the stress response changes from typical thixotropy to the peculiar combined recovery-and-decay dynamics, shows a concentration dependence. We are unable to precisely determine the value of the critical shear rate because of the large spacing between the shear rates applied. But in general, the higher the concentration, the larger the critical shear rate. This suggests that a larger hydrodynamic stress is required for more concentrated CB suspensions to undergo the structural change that is responsible for the two-regions stress response and the recovery and decay at low shear rates. The structural origins are discussed in Sec.~\ref{sec:structure}.

\begin{figure*}[h!]
	\begin{minipage}[ht]{0.49\textwidth}
		\centering
		\textbf{(a)}\par
		\includegraphics[scale=1,trim={0 0 0 0},clip]{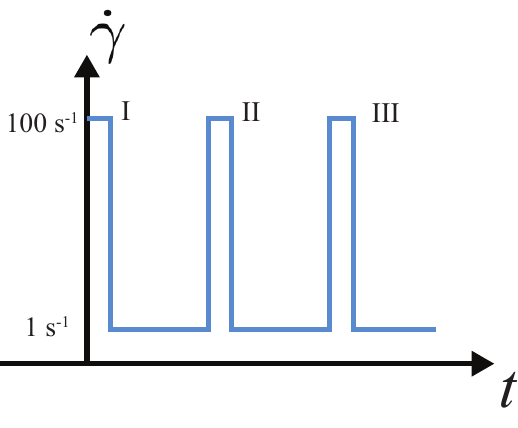}
	\end{minipage}
	\begin{minipage}[ht]{0.49\textwidth}
		\centering
		\textbf{(b)}\par
		\includegraphics[scale=0.4,trim={0 0 0 0},clip]{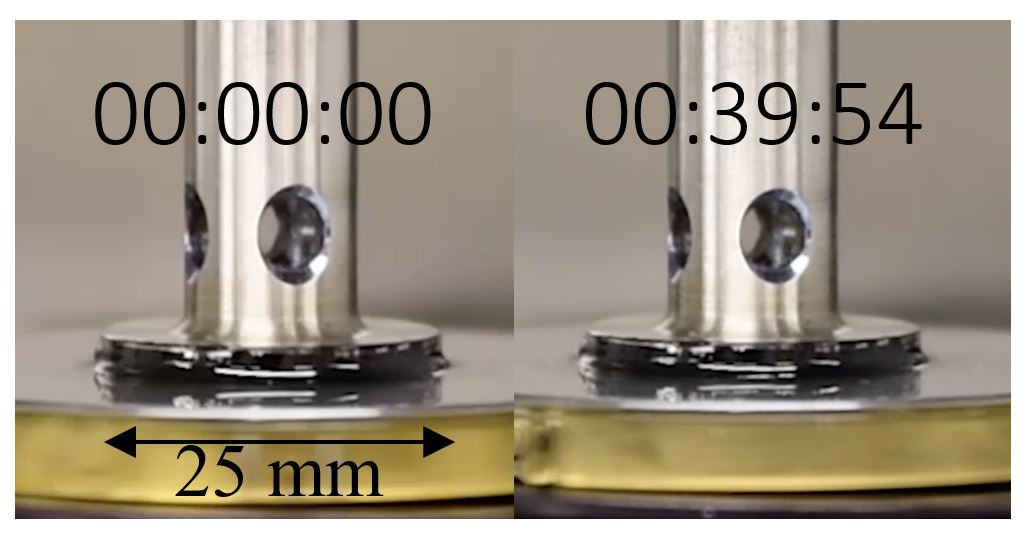}
	\end{minipage}
	\vskip20pt
	\begin{minipage}[ht]{0.49\textwidth}
	\centering
	\textbf{(c)}\par
	\includegraphics[scale=0.35,trim={0 0 0 0},clip]{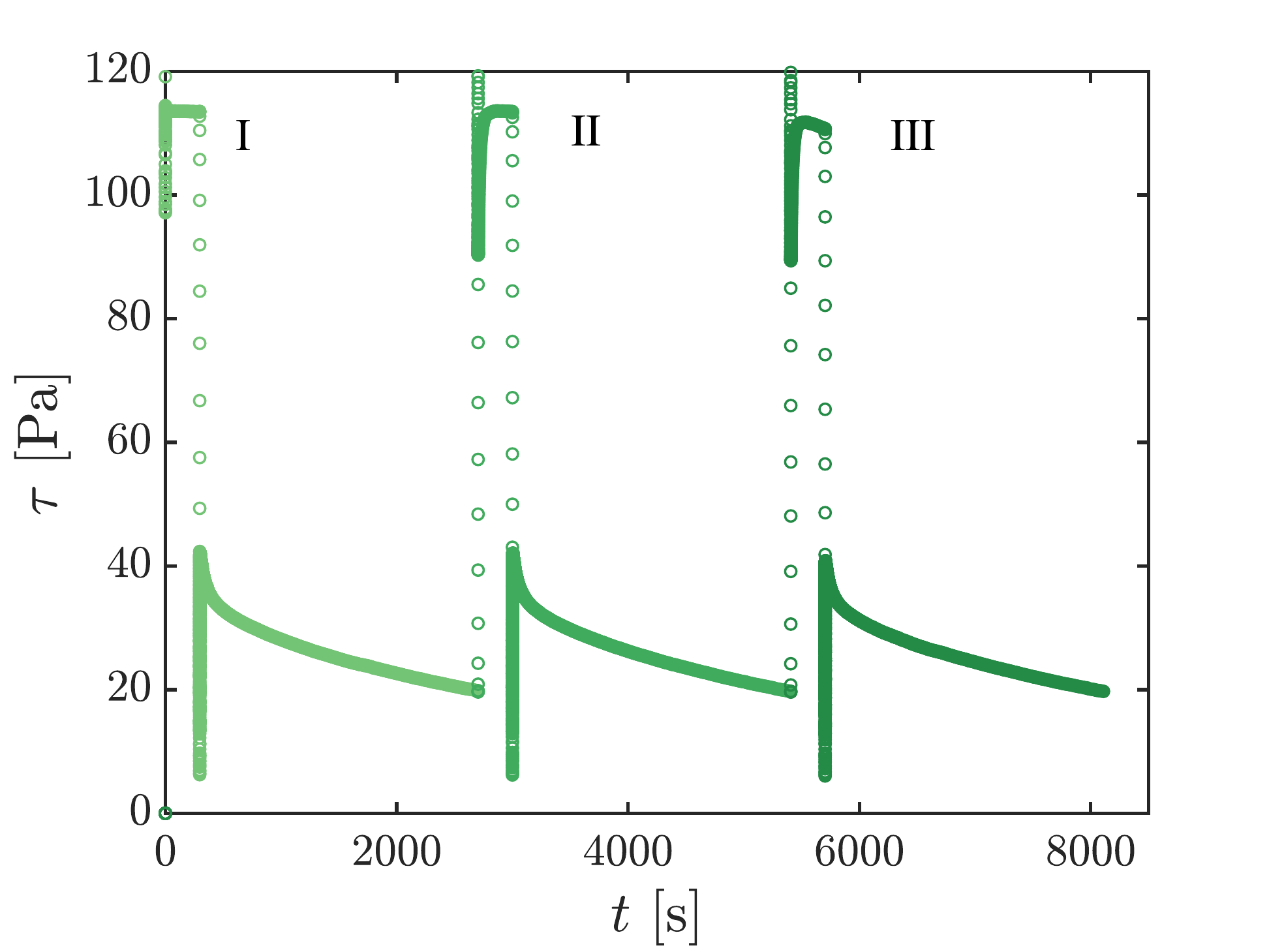}
	\end{minipage}
	\begin{minipage}[ht]{0.49\textwidth}
	\centering
	\textbf{(d)}\par
	\includegraphics[scale=0.45,trim={0 0 0 0},clip]{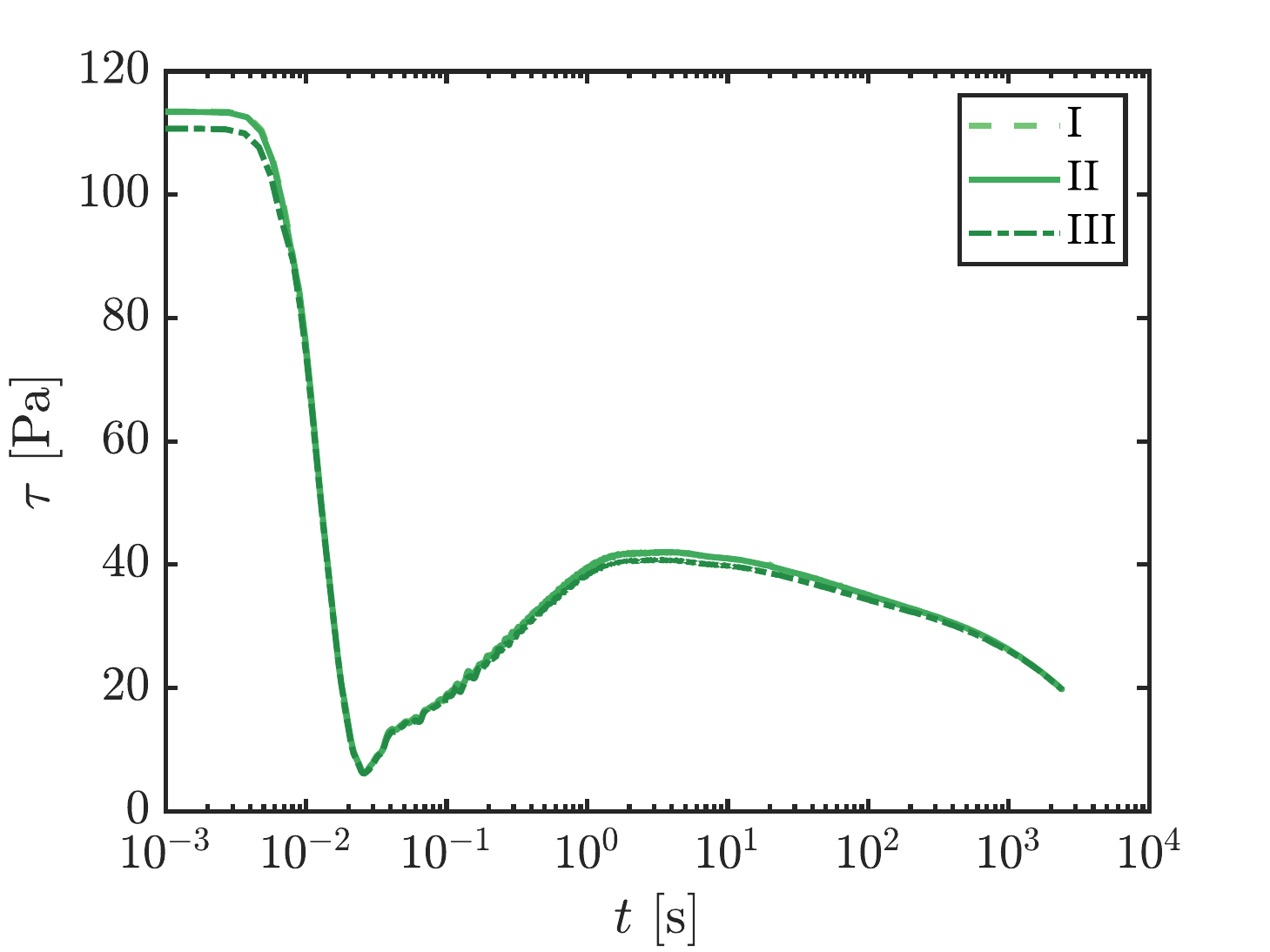}
	\end{minipage}
	\caption{Verifying the reversibility and repeatability of the shear stress response. The step shear as shown in (a) is applied on the 6~wt$\%$ CB suspension, and the edge of the sample is observed during the test, as shown in (b). The stress response is measured as (c) and is also plotted in logarithmic time scale as (d). The edge does not change during the test, ruling out artifacts such as edge failure. The repeated measurement results show that the recovery-and-decay dynamics is reversible and repeatable, as the stress responses for the three steps of time (I, II, and III) overlap.}
	\label{Fig.thixo_reversible}
\end{figure*} 

The reversibility and repeatability of the shear stress response was verified with a repeated step shear rate scheduling using a 25~$\rm mm$ diameter parallel plate geometry. Fig.~\ref{Fig.thixo_reversible}(a) shows the input shear rate, where the 6~wt$\%$ CB suspension is sheared at 100~$\rm s^{-1}$ for 120 seconds, then the rate is stepped down to 1~$\rm s^{-1}$ and maintained for 40 minutes. The shear rate is stepped up to 100~$\rm s^{-1}$ again and this cycle is repeated for three times without any time delay. In the meantime, the edge of the CB suspension sample is visually recorded. The video of the repeated test is available in the SI, and the photos of the CB sample before and after tests are shown in Fig. \ref{Fig.thixo_reversible}(b). From the recording we can see that the sample remains identical before and after the test, and there is no visual edge failure or sample loss during the whole process. This rules out experimental artifacts such as sample loss or edge failure. The stress response over time during the reversibility tests is plotted in Fig.~\ref{Fig.thixo_reversible}(c). In all three sets of step-down shear rate tests, the shear stresses show a sudden jump, followed by a decay at long times, not reaching the steady state even after 40 minutes. We further coplot the stress responses for each set of the three repeated tests in Fig.~\ref{Fig.thixo_reversible}(d) on a semilog scale, and the stress responses overlap with each other, which shows that the peculiar combined stress recovery and decay dynamics is reversible and repeatable.

\begin{figure}[h!]
	\includegraphics[scale=.6]{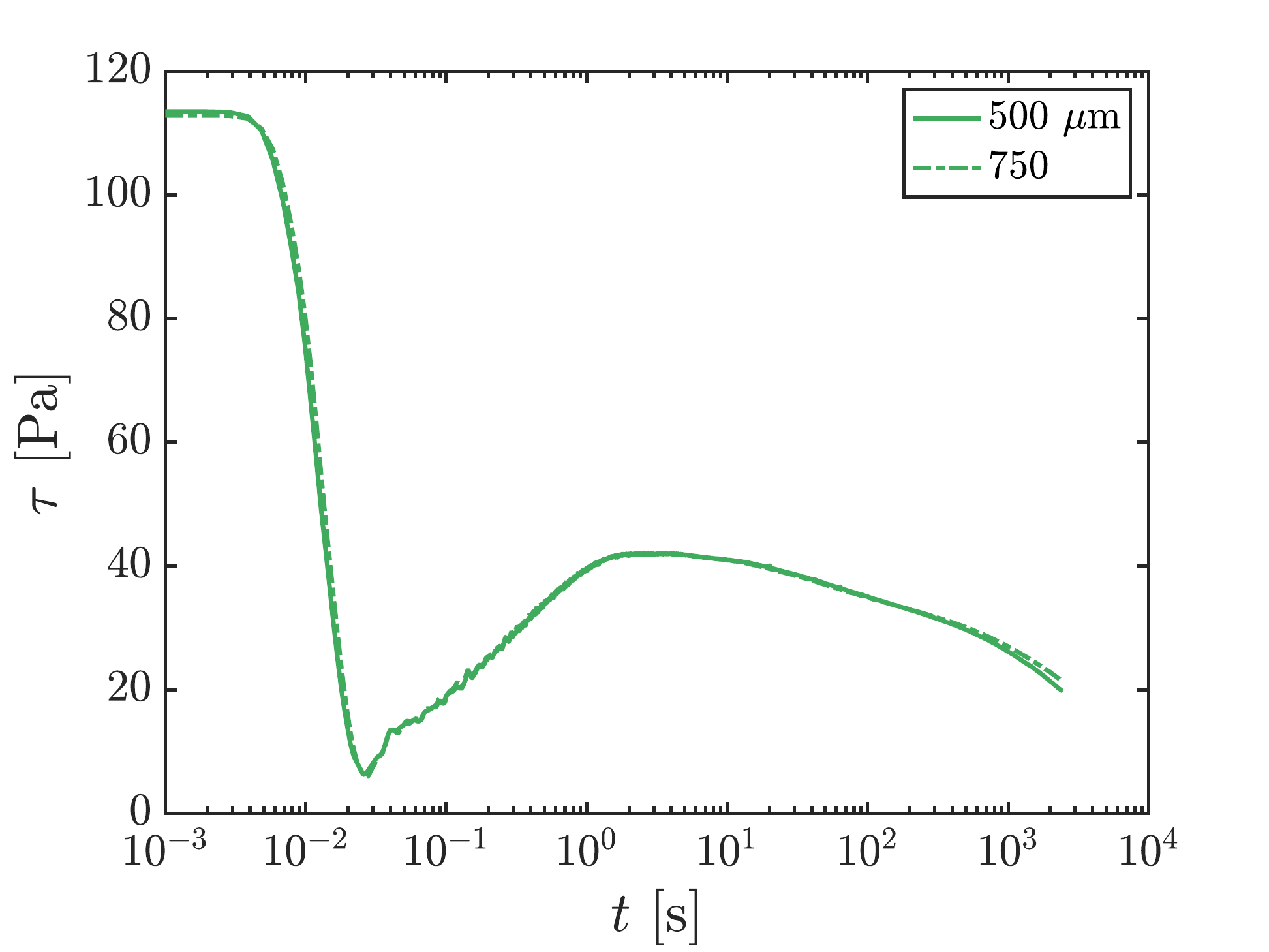}	
	\caption{Verifying the absence of wall-slip and shear-banding for the combined recovery-and-decay dynamics by measuring the transient stress response during the step-down in shear rate test at two different gaps using a 25~mm parallel plate geometry. The two stresses do not change with the measuring gap, showing that there is no slip or shear-banding at the shear rates tested.}
	\label{Fig.VulcanCB_6pctwt_pp_slip}
\end{figure}

Next, we verify that the dynamics is not due to wall-slip or shear banding, by confirming the reproducibility of the step-down shear rate tests at two different operating gaps, 500 and 750~$\rm \mu m$, using the 25~$\rm mm$ diameter parallel plate geometry \cite{Ewoldt2015}. Fig.~\ref{Fig.VulcanCB_6pctwt_pp_slip} shows the transient shear stress evolution, when the shear rate is decreased from 100~$\rm s^{-1}$ to 1~$\rm s^{-1}$, at different gaps. The stress responses do not depend on the operating gap, proving that there is no wall slip or shear banding occurring during the step-down shear rate tests for the shear rates tested. This is consistent with the results in literature, where wall slip is found to occur only at much lower shear rates (around $0.05 ~\rm  s^{-1}$ and lower)\cite{Dullaert2005}.  It should be noted that at short times, the stress signals oscillate around the mean value. The oscillation is not due to uneven sample loading or viscoelastic material waves, proved by a mismatch between stress oscillating frequency, motor rotating frequency, and the viscoelastic wave frequency, shown in the SI. Another possible reason for such oscillation is the force balancing transducer response, which we did not investigate here. Although the reason behind the oscillations is undetermined, the recovery trend at short times is distinct.

 We therefore conclude that the peculiar combined short-time recovery and long-time decay dynamics is a true material response, instead of being due to common experimental artifacts such as edge failure, slip, or shear banding, and it is repeatable and reversible. However, whether the decay at long times comes from anti-thixotropy or viscoelasticity remains unknown, as both anti-thixotropic and viscoelastic materials show stress decay in a step-down in shear test (Fig.~\ref{Fig.thixo_anti-thixo_VE}). In the next section, we will show that OSP combined with a step-down in shear rate test is needed to clarify the origin of the stress decay, from which we conclude that the decay is indeed anti-thixotropic and is related to shear-induced structure rearrangement.

\subsection{\label{sec:osp_results}Verification of anti-thixotropic structure rearrangement using OSP}
\begin{figure*}[ht]
	\begin{minipage}[]{0.49\textwidth}
		\centering
		\textbf{(a)}\par
		\includegraphics[scale=0.45,trim={0 0 0 0},clip]{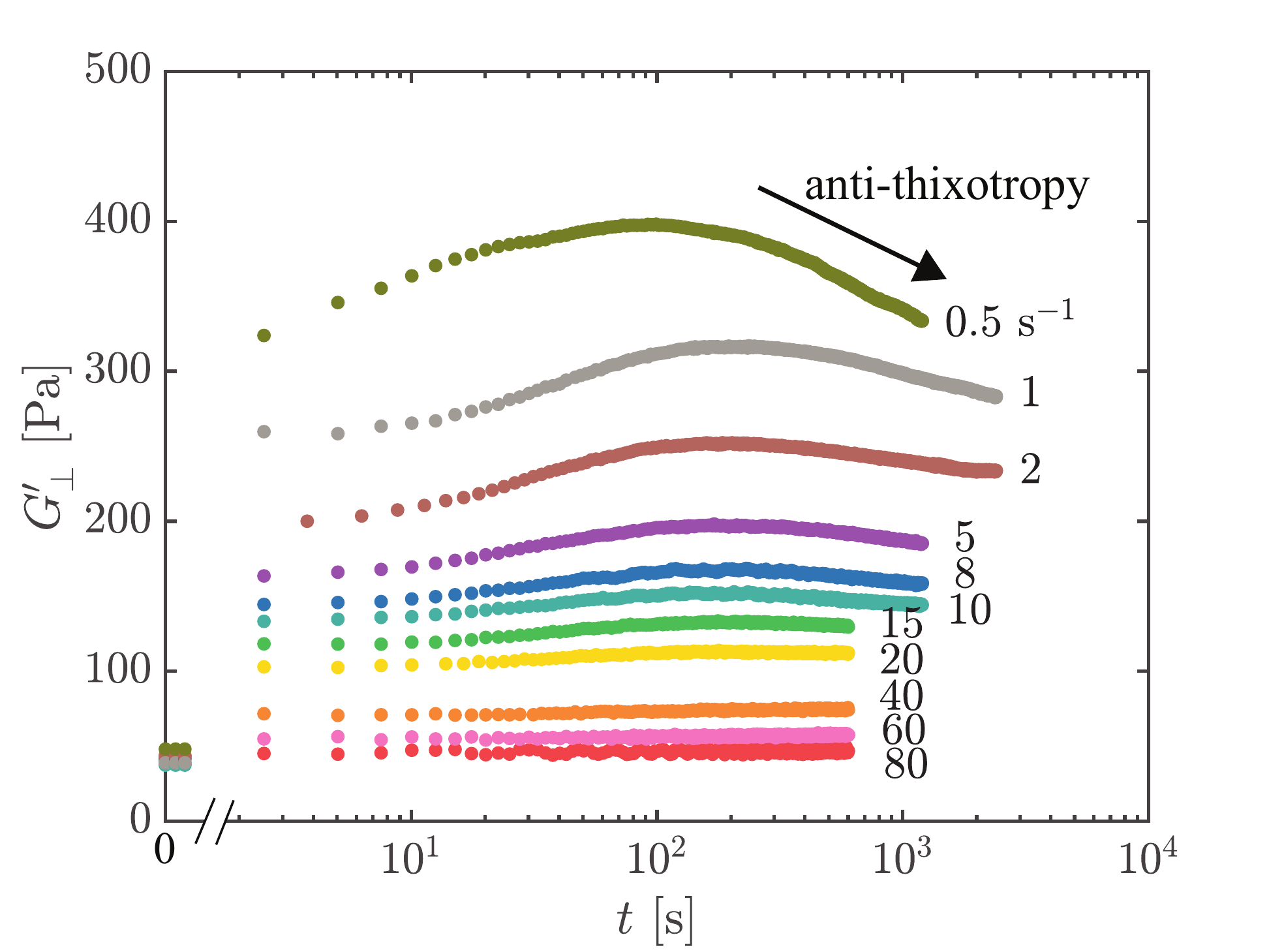}
	\end{minipage}
	\begin{minipage}[]{0.49\textwidth}
		\centering
		\textbf{(b)}\par
		\includegraphics[scale=0.45,trim={0 0 0 0},clip]{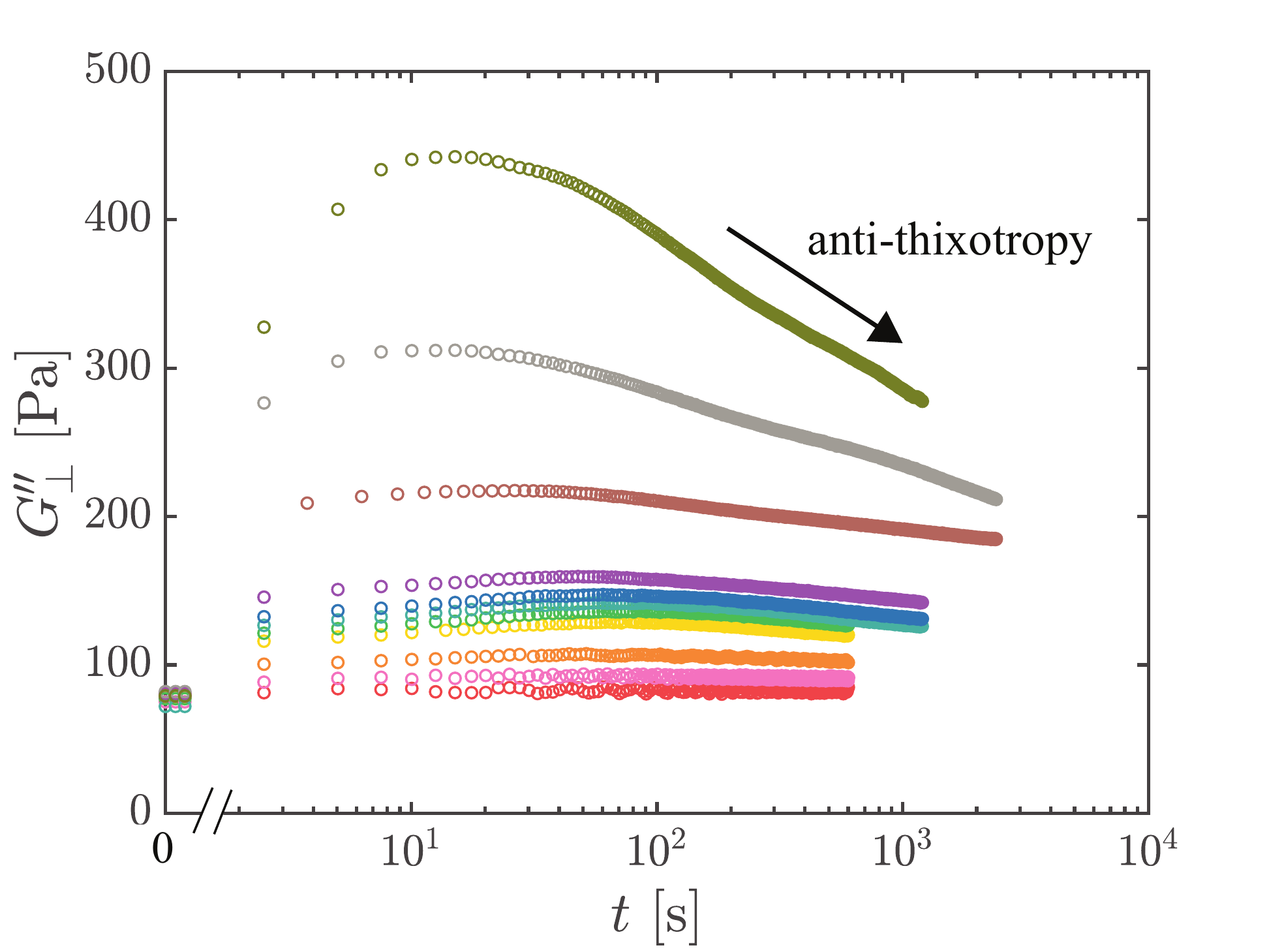}
	\end{minipage}
	\caption{The OSP results prove that the decay at long times comes from anti-thixotropy. The transient orthogonal (a) storage and (b) loss moduli of 6~wt$\%$ CB suspensions, measured by superposing an orthogonal oscillatory flow with 0.5$\%$ and 5~rad/s on the step-down shear rate flow in the rotational direction.}
	\label{Fig.VulcanCB_6pctwt_OSP_stepshear}
\end{figure*}

Orthogonal superposition (OSP) can potentially help to distinguish between anti-thixotropy and viscoelasticity, as responses from elastic and viscous components are separated. Fig.~\ref{Fig.VulcanCB_6pctwt_OSP_stepshear} shows the time evolution of orthogonal moduli, $G^{\prime}_\perp$ and $G^{\prime \prime}_\perp$, for the 6~wt$\%$ CB suspension, where the solid circles in Fig.~\ref{Fig.VulcanCB_6pctwt_OSP_stepshear}(a) are $G^{\prime}_\perp$, and open circles in  Fig.~\ref{Fig.VulcanCB_6pctwt_OSP_stepshear}(b) are $G^{\prime \prime}_\perp$. Similar to the transient shear stress responses under step shear tests, two distinct ranges in moduli responses are observed. In the high shear rate range, both $G^{\prime}_\perp$ and $G^{\prime \prime}_\perp$ recover from lower values because of the structure build-up, while in the low shear rate range, a combined recovery-and-decay dynamics appears in both storage and loss moduli. When $\dot \gamma_f $ is high, the recovery is very fast, and the moduli increase and reach steady states within 3~s. This response time is shorter than the time needed to collect oscillatory data at 5~rad/s, due to transients and multiple cycles averaging, and therefore the increase in $G^{\prime}_\perp$ and $G^{\prime \prime}_\perp$ cannot be resolved for times $\lesssim 3 \rm ~s$. With decreasing $\dot \gamma_f$, the recovery occurs over a longer timescale; the moduli reaching their respective maximum values after $\sim$~100~s for $G^{\prime}_\perp$ and $\sim$~10~s for $G^{\prime \prime}_\perp$. Beyond their respective maximum values, both $G^{\prime}_\perp$ and $G^{\prime \prime}_\perp$ start decreasing with no apparent steady state values even at extremely long times, identical to the stress transients in the low rate region. The decrease observed in both storage and loss moduli suggests that the long-time decay is related to structural rearrangement of CB, which changes both the elastic and viscous properties. Therefore, the long time decay observed is a result of anti-thixotropy, instead of viscoelastic stress relaxation. From Fig.~\ref{Fig.VulcanCB_6pctwt_OSP_stepshear} we can also see that the critical shear rate, below which the moduli start to show such combined dynamics, is around 15~$\rm s^{-1}$, although it cannot be determined precisely. The value of the critical shear rate is about the same as observed in the shear stress response in Fig.~\ref{Fig.VulcanCB_6pctwt_stepshear_stress}. This suggests that the transient responses of shear stress and orthogonal moduli have the same structural origin, which is a result of CB structure rearrangement involving not only viscous, but also elastic stress contributions.

\begin{figure}[h!]
	\includegraphics[scale=.6]{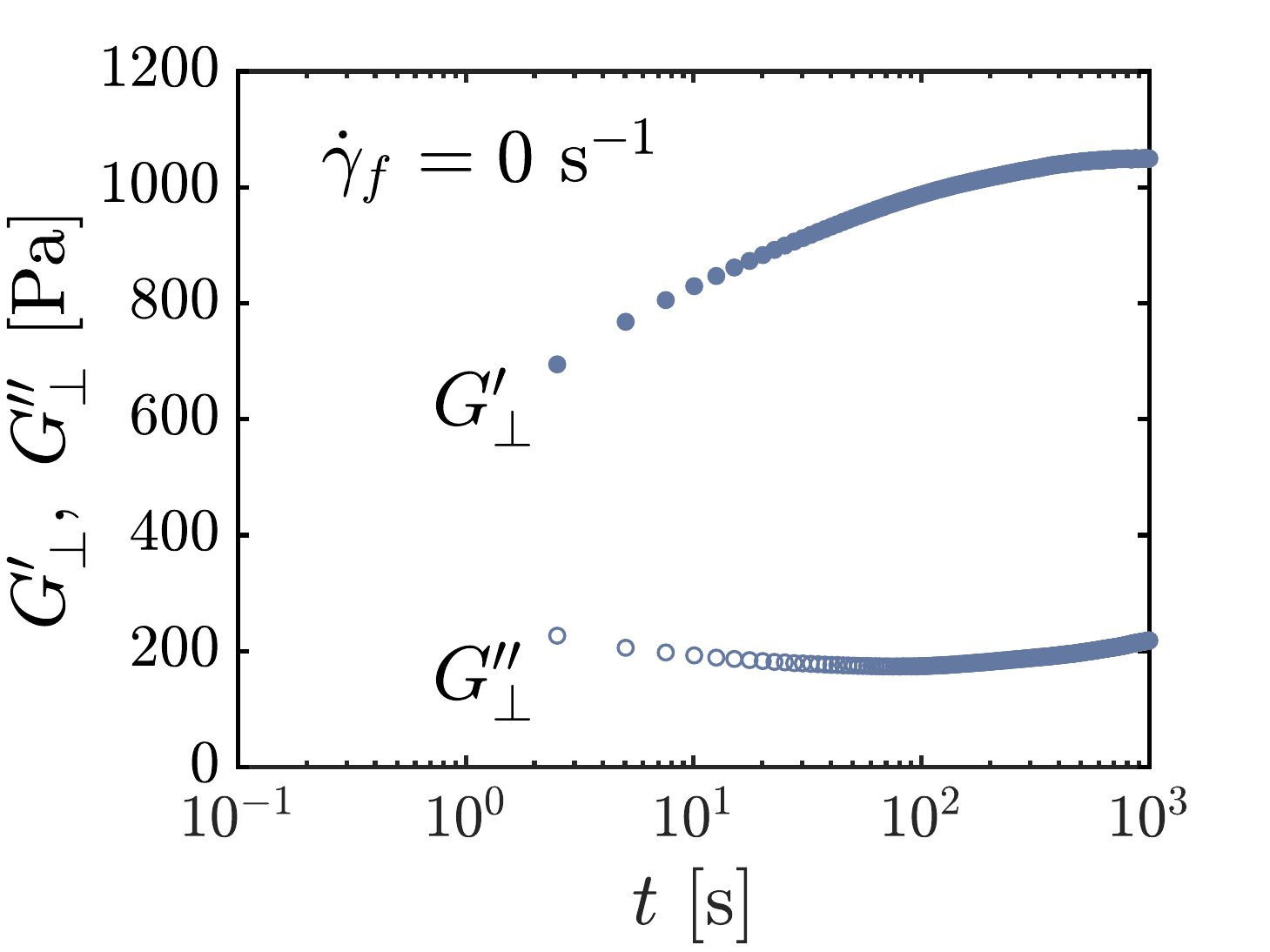}	
	\caption{The results of cessation of steady shear flow at 100 $\rm s^{-1}$ show no long-time decay, and therefore such decay as in Fig \ref{Fig.VulcanCB_6pctwt_OSP_stepshear} and Fig. \ref{Fig.OSP_reversible} is shear-induced, which again verifies that the decay is anti-thixotropic. The transient orthogonal storage modulus (solid circle) and loss modulus (open circle) of a 6 wt$\%$ CB suspension measured during the flow cessation test, where the flow is stopped from shearing at the initial shear rate of 100 $\rm s^{-1}$, while measuring the orthogonal moduli using OSP simultaneously throughout.}
	\label{Fig.VulcanCB_6pctwt_stepshear_coplot_100_0}
\end{figure}

The long-time decay is shown to be shear-induced with the help of OSP, providing additional evidence for anti-thixotropic structure rearrangements rather than viscoelastic relaxation. Fig.~\ref{Fig.VulcanCB_6pctwt_stepshear_coplot_100_0} shows $G^{\prime}_\perp(t)$ and $G^{\prime \prime}_\perp(t)$ during the cessation of steady flow at 100~$\rm s^{-1}$. Without applied shear, the orthogonal moduli only show a thixotropic recovery, increasing monotonically to their respective steady state values, and the storage modulus is much larger than the loss modulus during the cessation. The increase in storage modulus suggests that the suspension undergoes elastic stiffening at rest, rather than the structure rearrangement that decreases the viscosity and moduli. The increase in storage modulus at rest is consistent with the results on fumed silica suspensions \cite{Colombo2017} and has been reported on CB suspensions \cite{Osuji2008a}, where the value of the storage modulus depends on the applied shear stress before cessation. The stiffening at rest also suggests that a low but definite applied shear rate is required for CB suspensions to undergo the structure rearrangement that decreases the viscosity and moduli, and the decay in both the viscosity and moduli is therefore shear-induced. This provides overwhelming evidence that the decrease is not a result of viscoelastic relaxation, but anti-thixotropic shear-induced structural change. The ability to measure rheological properties changing with time after cessation of flow is another advantage of superposition rheometry, including orthogonal and parallel superposition, which is not possible with the standard cessation of shear because of no shear applied. The same technique, using OSP during cessation of steady flow, has been used on fumed silica suspensions to study the anisotropy of the resulting colloidal gel \cite{Colombo2017}, but this is the first time OSP has been used on CB suspensions to study the anti-thixotropic shear-induced structuring of CB particles.

\begin{figure*}[ht]
	\begin{minipage}[!h]{0.3\textwidth}
		\centering
		\textbf{(a)}\par
		\includegraphics[scale=0.8,trim={0 0 0 0},clip]{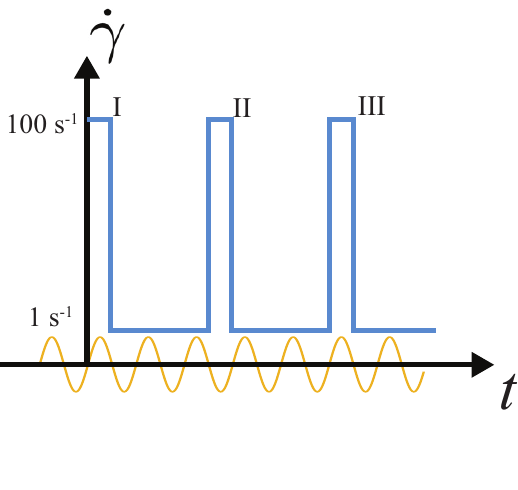}
	\end{minipage}
	\begin{minipage}[!h]{0.3\textwidth}
		\centering
		\textbf{(b)}\par
		\includegraphics[scale=0.28,trim={0 0 0 0},clip]{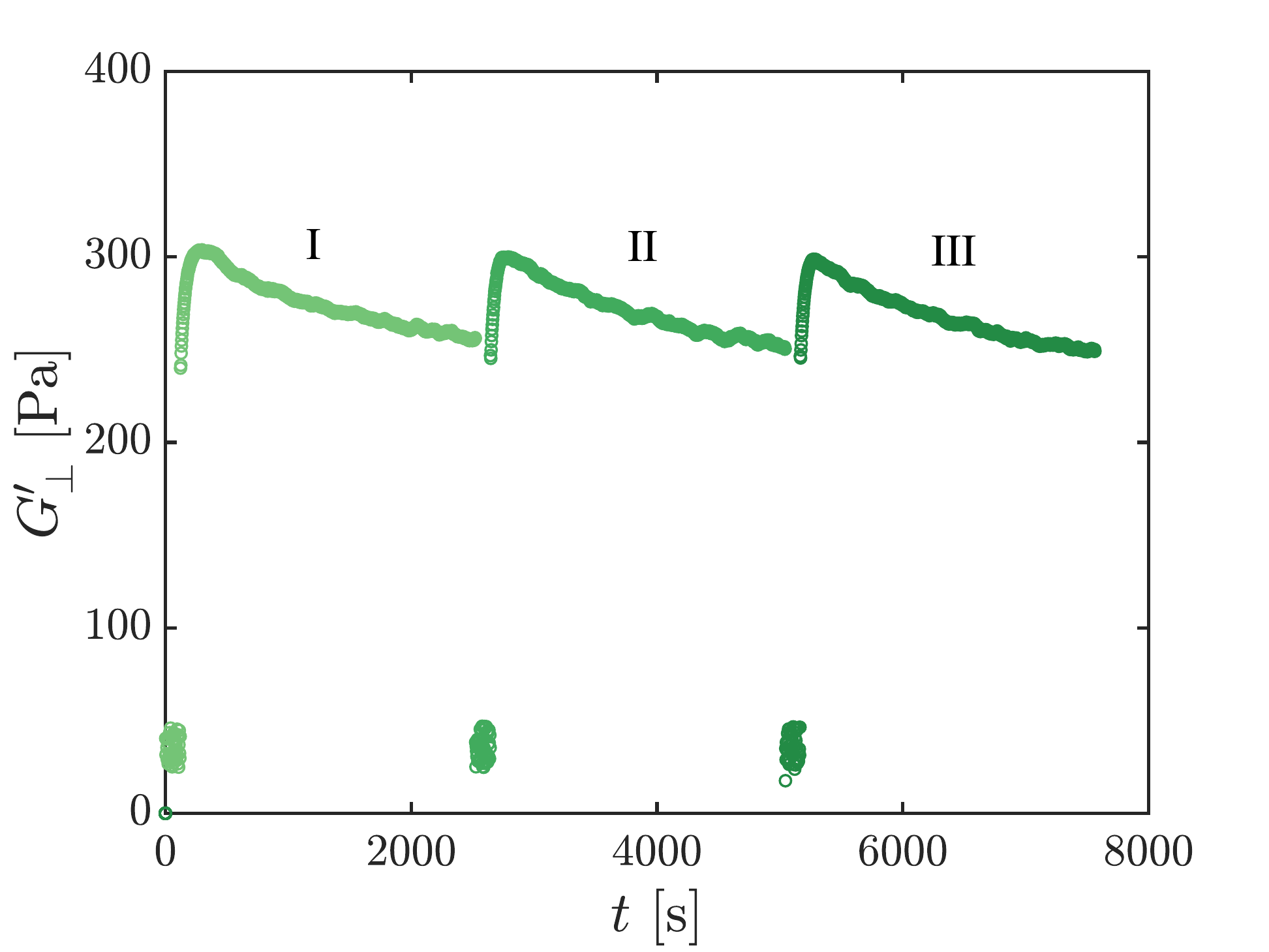}
	\end{minipage}
	\begin{minipage}[!h]{0.3\textwidth}
		\centering
		\textbf{(c)}\par
		\includegraphics[scale=0.28,trim={0 0 0 0},clip]{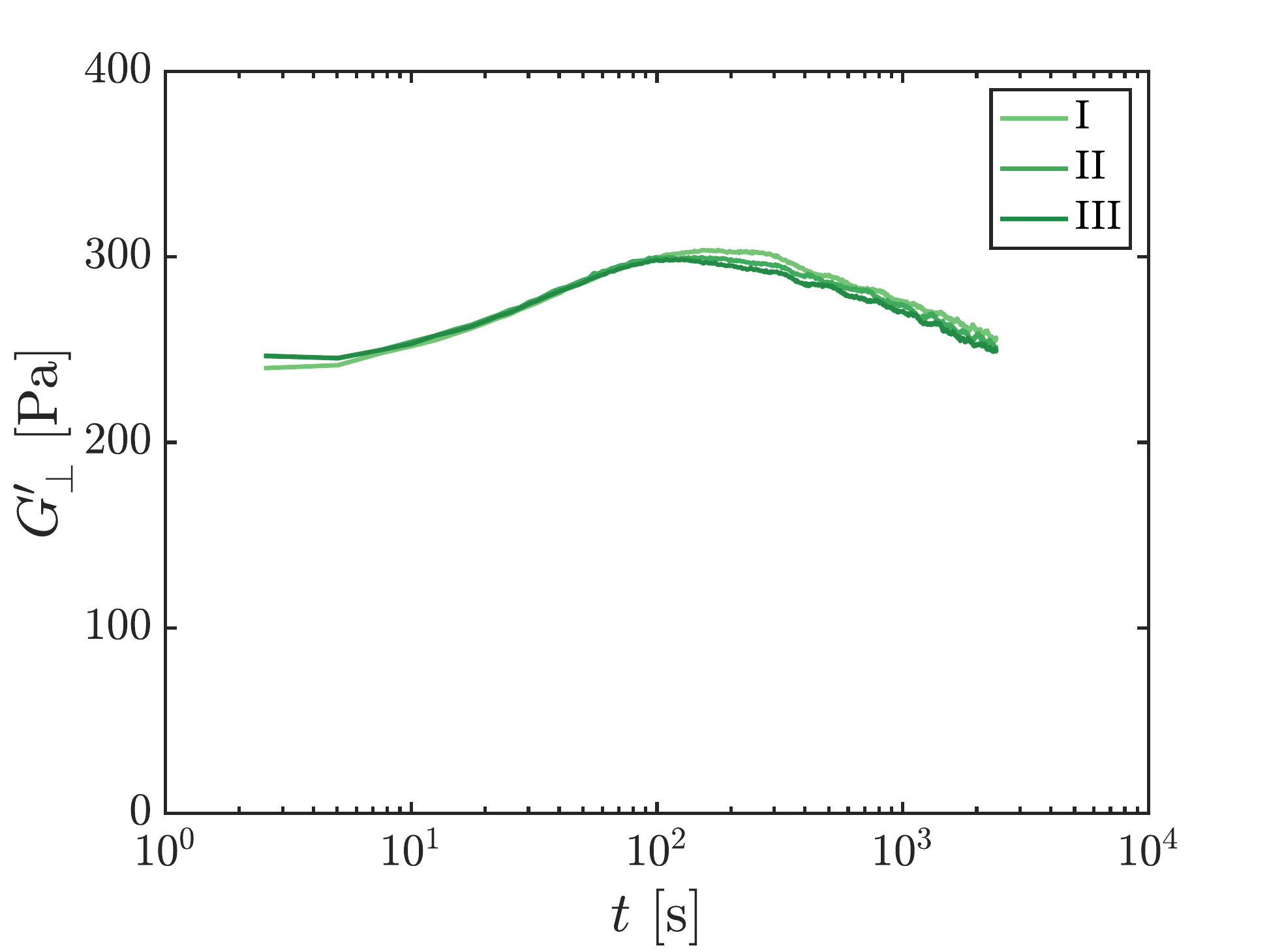}
	\end{minipage}
	\caption{Verifying the repeatability and reversibility of the orthogonal moduli response. The step shear rate flow and an oscillatory flow in the orthogonal direction with 5~rad/s and $0.5\%$ strain, as shown in (a) is applied on the 6~wt$\%$ CB suspension, and the storage modulus in the orthogonal direction is recorded, as shown in (b); the modulus response during the stepping-down for the three sets is also plotted on a log time scale, which shows the dynamics are repeatable and reversible, as the modulus responses for the three sets overlap.}
	\label{Fig.OSP_reversible}
\end{figure*}

The reversibility and repeatability of the OSP results are verified by applying the repeated step shear rate scheduling from 100 to 1~$\rm s^{-1}$ as shown in Fig.~\ref{Fig.OSP_reversible}(a), while measuring $G^{\prime}_\perp$ and $G^{\prime \prime}_\perp$ at $\omega=5$~rad/s and $\gamma_0 = 0.5\%$ simultaneously throughout. The corresponding storage modulus as a function of time during the repeated measurements is shown as Fig.~\ref{Fig.OSP_reversible}(b) and the moduli responses during the three sets of stepping-down tests are co-plotted on a logarithmic time scale in Fig.~\ref{Fig.OSP_reversible}(c). The loss modulus is omitted for simplicity. The response is reversible in repeated tests and the storage moduli overlap (Fig.~\ref{Fig.OSP_reversible}(c)), all showing the short-time thixotropic recovery followed by the long-time anti-thixotropic decay. This verifies that the OSP results are repeatable and reversible, and they come from a true material response, rather than experimental artifacts.

It should be noted that the fluctuation of both $G_\perp^\prime$ and $G_\perp^{\prime \prime}$, especially at the high shear rates, exists in both storage and loss moduli, which is likely to originate from Taylor vortices due to high angular rotation rates. Studying the limits of the operating windows of the OSP, however, is outside the scope of this work. Another caveat of using OSP is that the orthogonal oscillation can significantly disrupt particle contacts for some systems \cite{Lin2016}, and therefore change the properties in the primary shear direction. To avoid structure disruption by the orthogonal oscillation, the amplitude used here is as low as possible while being sufficient to generate a detectable stress signal. For the 6~wt$\%$ CB suspension, the amplitude is 0.5$\%$ and the frequency is 5~rad/s, which is in the linear viscoelastic regime of the suspension. The amplitude sweep results are shown in Fig.~S5. We also test the effect of orthogonal oscillation on the primary shear stress by comparing the shear stress in the rotational direction during OSP tests with the transient stress response shown in Fig.~\ref{Fig.VulcanCB_6pctwt_stepshear_stress}, which shows the same response at the same timescale, as shown in Fig.~S7. This suggests that the rheological properties, and therefore, the structure of CB in the primary flow direction are unaffected by the small amplitude orthogonal oscillation. 

The transient orthogonal moduli for CB suspensions of other weight fractions are shown in Fig.~S8-S10, which show that the thixotropic recovery and anti-thixotropic decay in the low final shear rate region are ubiquitous to all concentrations tested. OSP provides us with the first definitive evidence of the distinction between viscoelasticity and anti-thixotropy of CB suspensions, which is also the first instance of using OSP to resolve the ambiguity between viscoelasticity and anti-thixotropy. With the help of step-down in shear rate tests coupled with OSP, we conclude that the time varying mechanical response of the CB suspension under step shear is a combined short-time thixotropic recovery and long-time anti-thixotropic decay.

\subsection{\label{sec:timescale}Anisotropy in thixotropic timescales and moduli}

\begin{figure*}[ht]
	\begin{minipage}[!h]{0.3\textwidth}
		\centering
		\textbf{(a)}\par
		\includegraphics[scale=0.27,trim={0 0 0 0},clip]{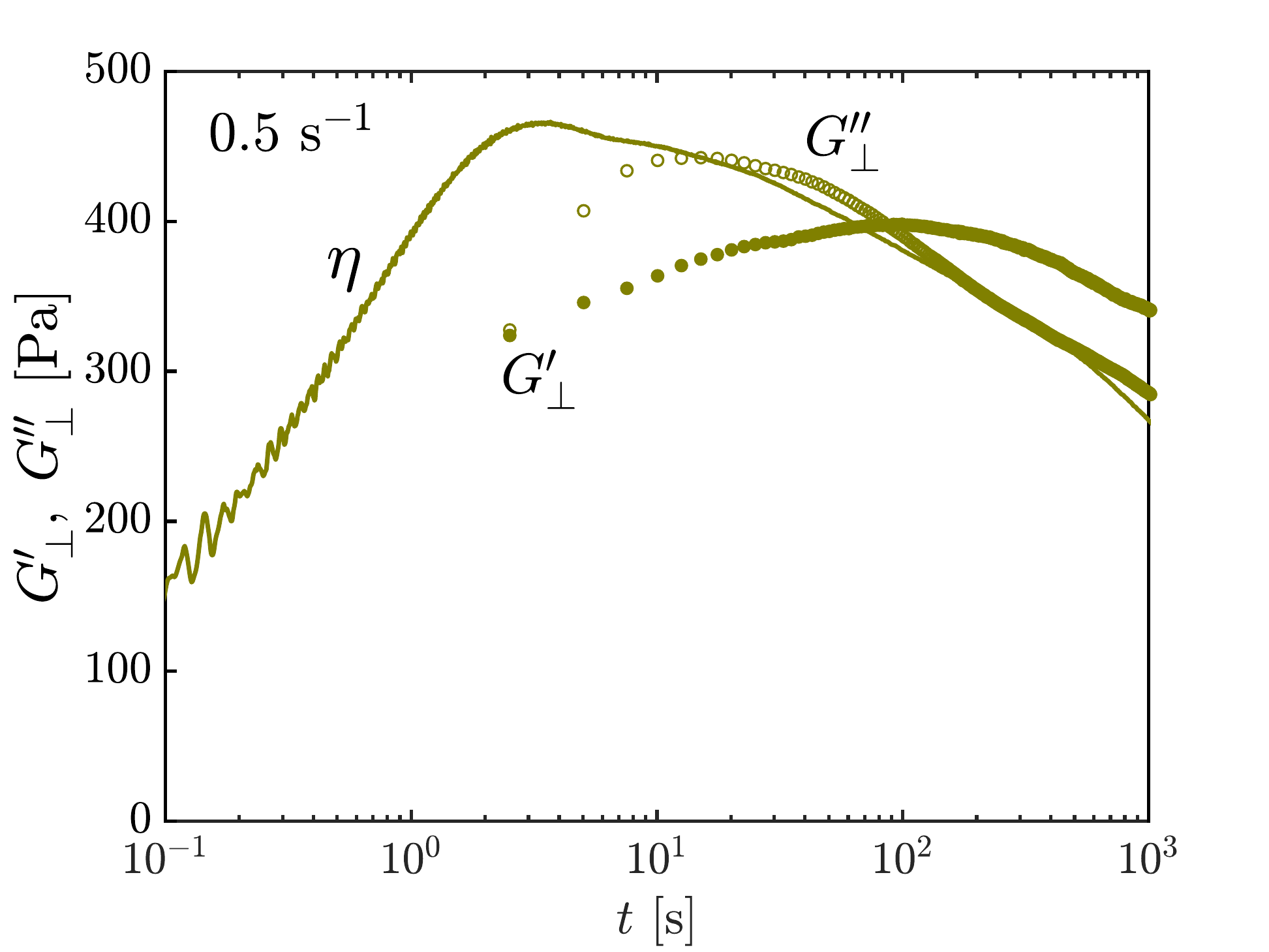}
	\end{minipage}
	\begin{minipage}[!h]{0.3\textwidth}
	\centering
	\textbf{(b)}\par
	\includegraphics[scale=0.27,trim={0 0 0 0},clip]{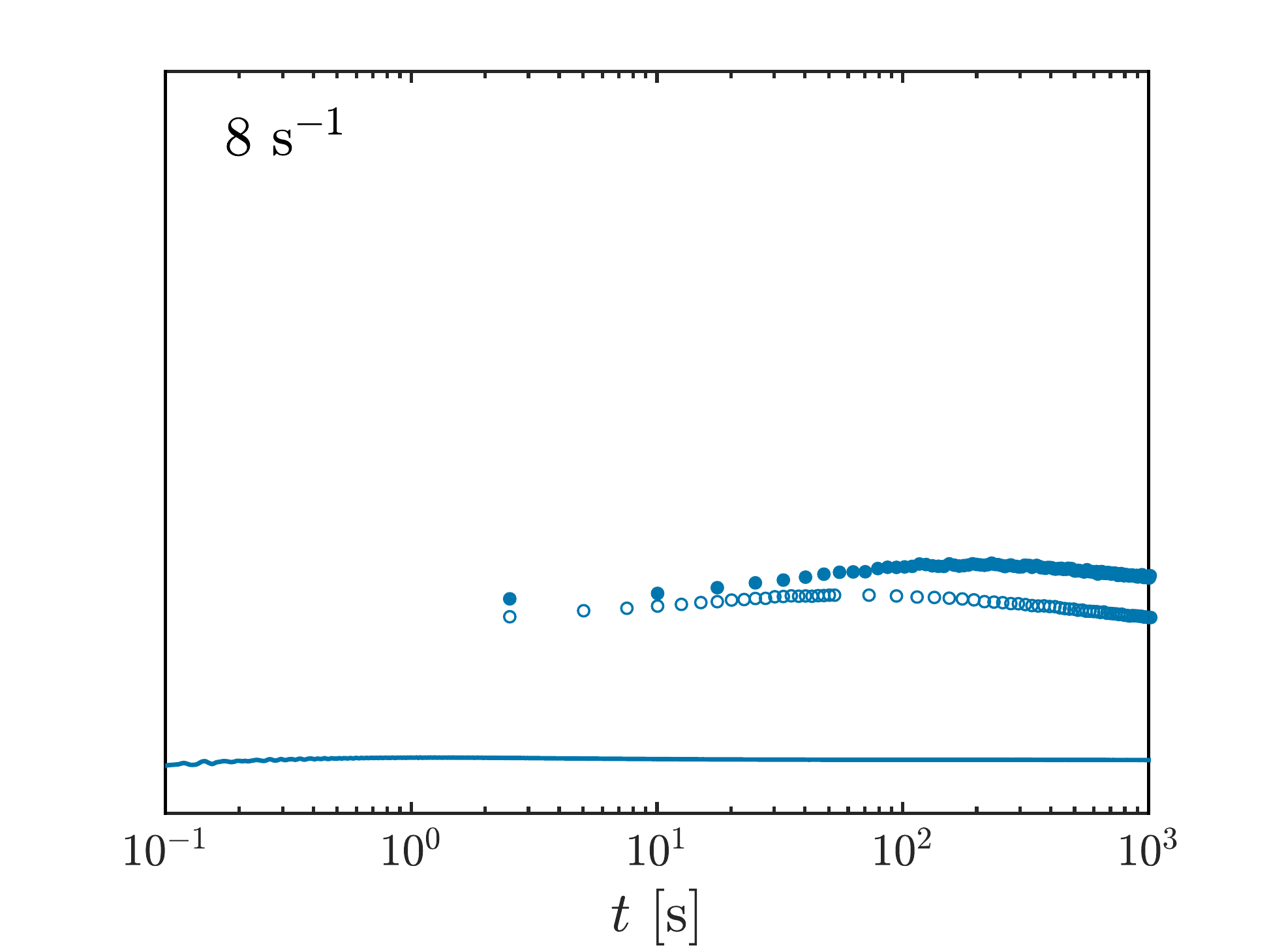}
	\end{minipage}
	\begin{minipage}[!h]{0.3\textwidth}
	\centering
	\textbf{(c)}\par
	\includegraphics[scale=0.27,trim={0 0 0 0},clip]{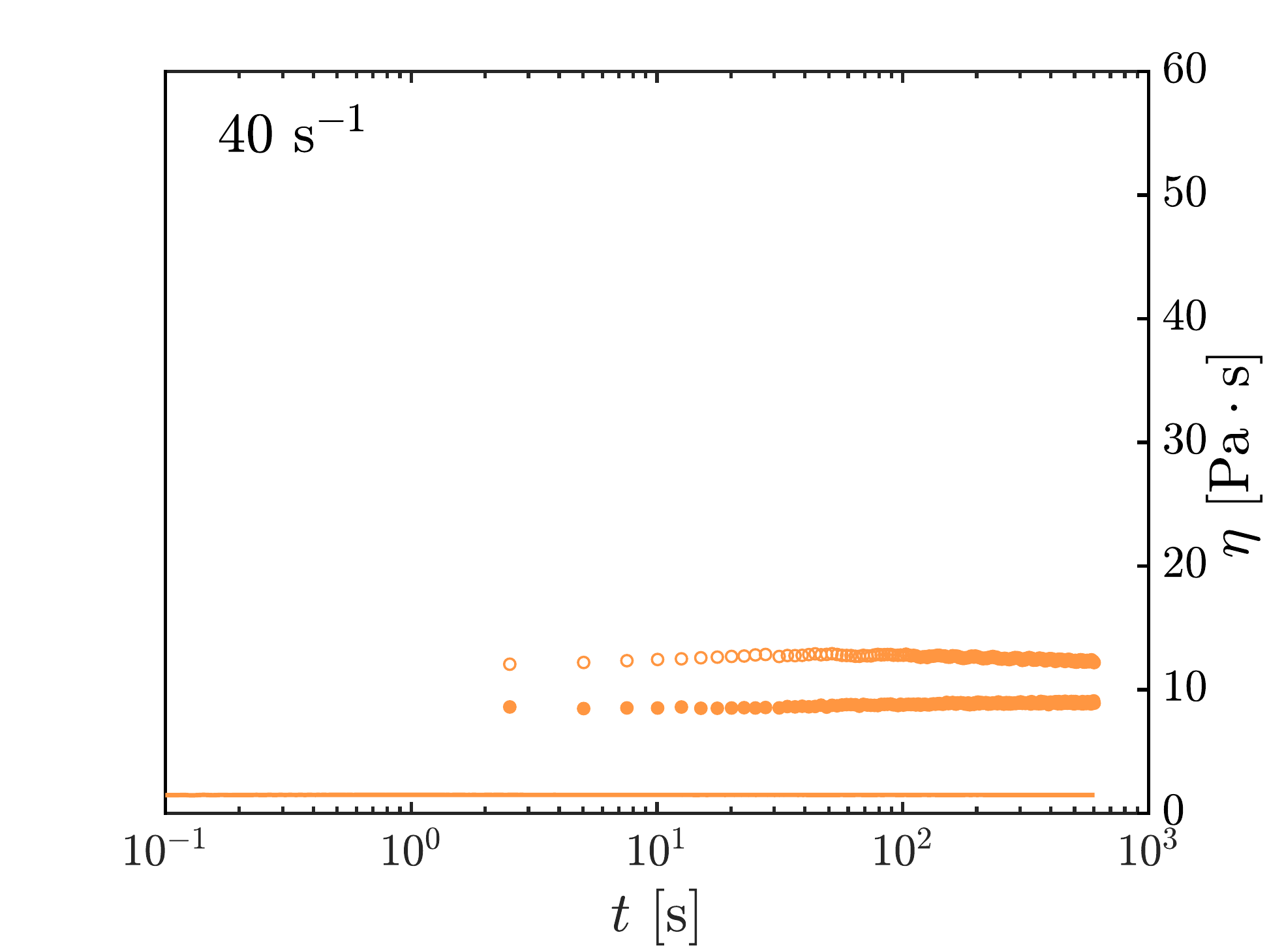}
	\end{minipage}
	\caption{Evidence of anisotropic dynamics from co-plotted parallel transient shear viscosity and orthogonal storage and viscous moduli during the step–down shear tests (data from Fig.~\ref{Fig.VulcanCB_6pctwt_stepshear_stress} and Fig.~\ref{Fig.VulcanCB_6pctwt_OSP_stepshear}). The shear rate is decreased from 100 to (a) 0.5; (b) 8; and (c) 40~$\rm s^{-1}$ respectively.}
	\label{Fig.VulcanCB_stepshear_coplot}
\end{figure*}

 We observe different timescales of thixotropic dynamics for different rheological measures, which indicates anisotropy in CB suspensions. Fig.~\ref{Fig.VulcanCB_stepshear_coplot} shows the orthogonal moduli, $G^{\prime}_\perp$ and $G^{\prime \prime}_\perp$, and shear viscosity, $\eta$, calculated from the shear stress, co-plotted for several final shear rates, 0.5, 8, and 40~$\rm s^{-1}$ (from data in Fig.~\ref{Fig.VulcanCB_6pctwt_stepshear_stress} and Fig.~\ref{Fig.VulcanCB_6pctwt_OSP_stepshear}). Consistent with earlier plots, the shear viscosity is plotted as a solid line, the orthogonal storage and loss moduli are plotted as solid and open circles respectively. It can be seen in Fig.~\ref{Fig.VulcanCB_stepshear_coplot} that at the same shear rate, the thixotropic timescales are different for shear viscosity in the rotational direction and moduli in the orthogonal direction. For example, when the final shear rate is $0.5~\rm s^{-1}$, as shown in Fig.~\ref{Fig.VulcanCB_stepshear_coplot}(a), the shear viscosity reaches its maximum first (at 3~s), followed by the orthogonal loss modulus (at 15~s), and finally the orthogonal storage modulus (at 100~s), taking more than one order of magnitude longer than the shear viscosity. The same trend is observed for $ \dot \gamma_f = 8 ~\rm s^{-1}$ (shown in Fig.~\ref{Fig.VulcanCB_stepshear_coplot}(b)) and $\dot \gamma_f = 40~\rm s^{-1}$ (Fig.~\ref{Fig.VulcanCB_stepshear_coplot}(c)). For all the shear rates tested ranging from 0.5 to 80~$\rm s^{-1}$, there is a separation of thixotropic timescales between shear stress/viscosity in the rotational direction and the moduli in the orthogonal direction: the anti-thixotropic structural rearrangement under shear is delayed in the orthogonal direction. The separation of thixotropic timescales suggests an anisotropic CB structure in the rotational and orthogonal directions, $e.g.$, due to the size of CB agglomerates longer in the orthogonal direction than in the rotational one.

\begin{figure*}[ht]
	\begin{minipage}[!h]{0.49\textwidth}
		\centering
		\textbf{(a)}\par
		\includegraphics[scale=0.43,trim={0 0 0 0},clip]{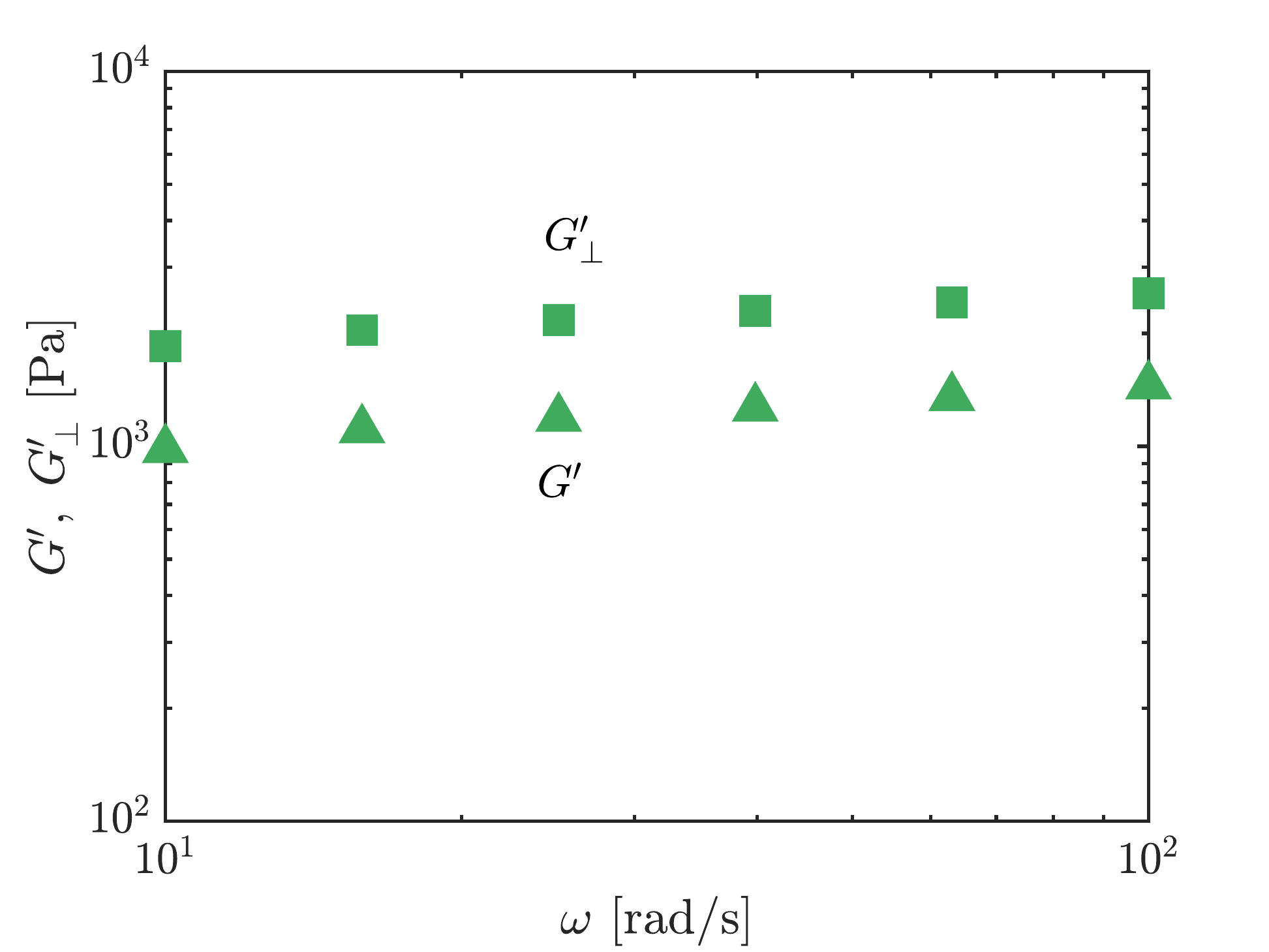}
	\end{minipage}
	\begin{minipage}[!h]{0.49\textwidth}
		\centering
		\textbf{(b)}\par
		\includegraphics[scale=0.43,trim={0 0 0 0},clip]{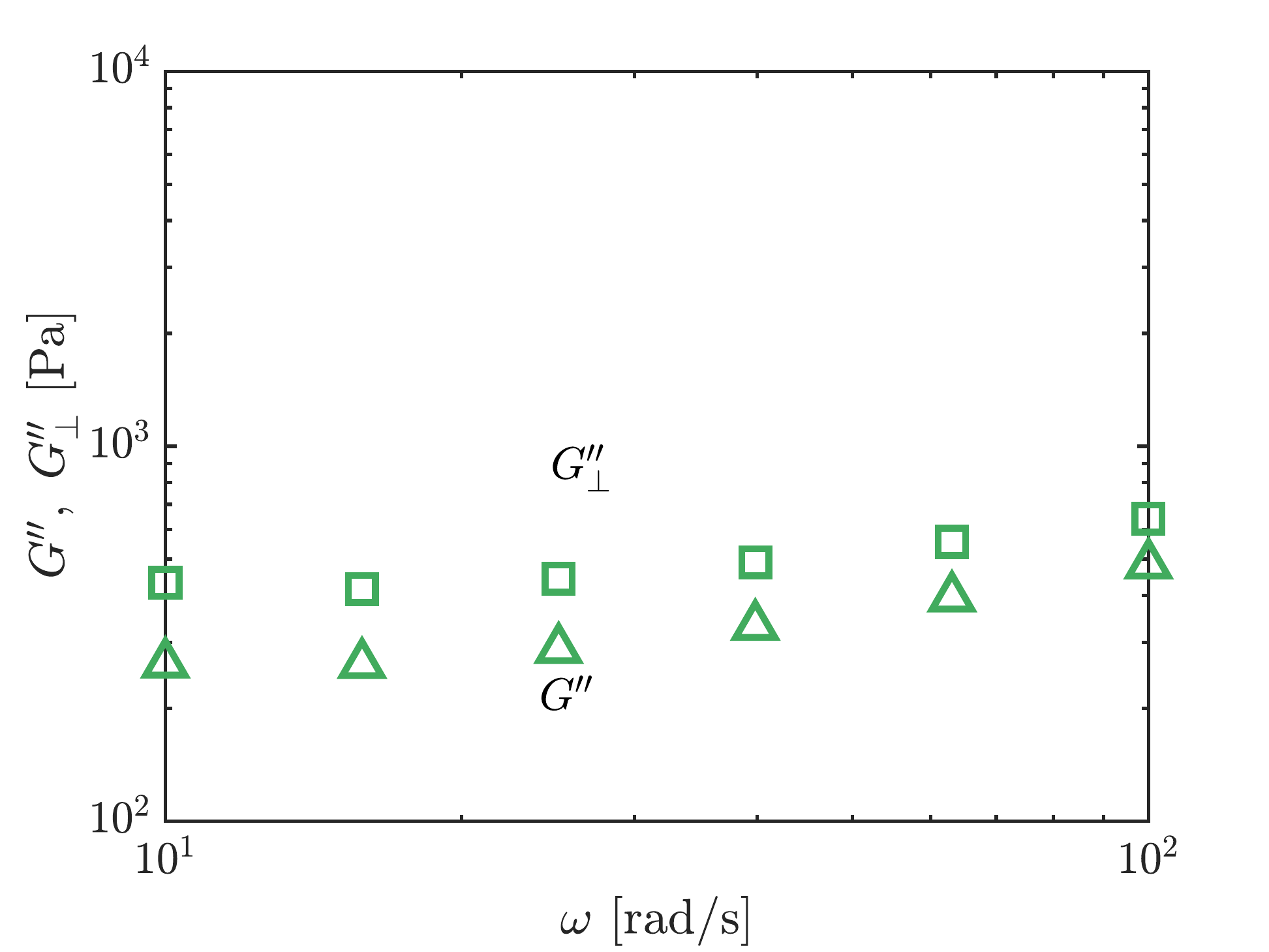}
	\end{minipage}
	\caption{Stiffer in the perpendicular direction. The anisotropic (a) elastic and (b) loss moduli of 6~wt$\%$ CB suspension for frequencies from 10 to 100~rad/s after shearing at 1~$\rm s^{-1}$ for 100~s, measured using OSP 2D oscillation with peak amplitude of 0.5~$\%$.}
	\label{Fig.VulcanCB_6pctwt_OSP_SAOS}
\end{figure*}

The mechanical anisotropy also shows in the difference between the amplitude of moduli, where the storage and loss moduli in the orthogonal direction are larger than those in the rotational direction by a factor of two. Fig.~\ref{Fig.VulcanCB_6pctwt_OSP_SAOS}(a) shows the storage moduli in the rotational direction, $G^\prime$, and those in the orthogonal direction, $G^{\prime}_\perp$, at frequencies ranging from 10 to 100 rad/s, measured using a 2D-small amplitude oscillatory shear (2D-SAOS) at a peak amplitude of $0.5\%$, while Fig.~\ref{Fig.VulcanCB_6pctwt_OSP_SAOS}(b) shows the loss moduli, $G^{\prime \prime}$ and $G^{\prime \prime}_\perp$. Both storage and loss moduli are larger in the orthogonal direction than those in the rotational direction. As a benchmark test, the moduli in both directions are measured using the same OSP geometry on aqueous Carbopol 940 at 1~wt$\%$, which is known as a homogeneous, isotropic, model yield stress material, and our results show that both storage and loss moduli are the same in both directions within 10$\%$; details in Fig.~S11. This validates the anisotropic moduli measured on CB suspensions to be of a structural origin, and not experimental artifacts. We can also see from Fig.~\ref{Fig.VulcanCB_6pctwt_OSP_SAOS} that all moduli show a weak dependence on the frequency, and storage moduli are larger than the loss moduli, which are signatures of soft solid behavior. Similar trends in the moduli can be observed for CB suspensions at the other concentrations tested, which suggests a structural anisotropy across all concentrations tested. The mechanical anisotropy has also been shown using OSP on other attractive colloidal systems such as fumed silica suspensions, where the storage moduli in the orthogonal direction are larger\cite{Colombo2017} than the rotational storage moduli by a factor of two to a hundred, depending on the shear conditions.

The mechanical anisotropy in thixotropic timescales and the magnitude of linear moduli suggests the formation of anisotropic CB agglomerates under the applied low shear rate. We hypothesize that this arises from long slender structures aligned in the orthogonal vorticity direction. Such "log rolling" anisotropic structures for CB suspensions have been reported before \cite{Varga2019}. It is shear-induced, formed at low shear rates, aligned in the vorticity (orthogonal, in this case) direction, and separated from each other by a certain distance to form a striped pattern. The anisotropic log-rolling flocs have been observed in the same type of CB suspensions of various concentrations using rheo-optical setups \cite{Grenard2011, Osuji2008}. The striped pattern of the log-rolling structure is only realized in the range of 1-2~wt$\%$; at the higher concentrations (2.5-3~wt$\%$) only several "holes" are observable between the large agglomerates \cite{Grenard2011}. At concentrations tested here (6-10~wt$\%$), the log-rolling structure is not visible using optical setups, likely due to the high CB density and an overlap between large flocs. However, the mechanical anisotropy in both the thixotropic timescale and the moduli measured using OSP are consistent with the hypothesis of anisotropic vorticity-aligned structures. The rheological OSP technique is robust and versatile in detecting the anisotropic structure even without high-resolution microscopy.

\subsection{\label{sec:concentration}Concentration dependence of stresses and moduli}

\begin{figure*}[ht]
	\begin{minipage}[!h]{0.3\textwidth}
		\centering
		\textbf{(a)}\par
		\includegraphics[scale=0.27,trim={0 0 0 0},clip]{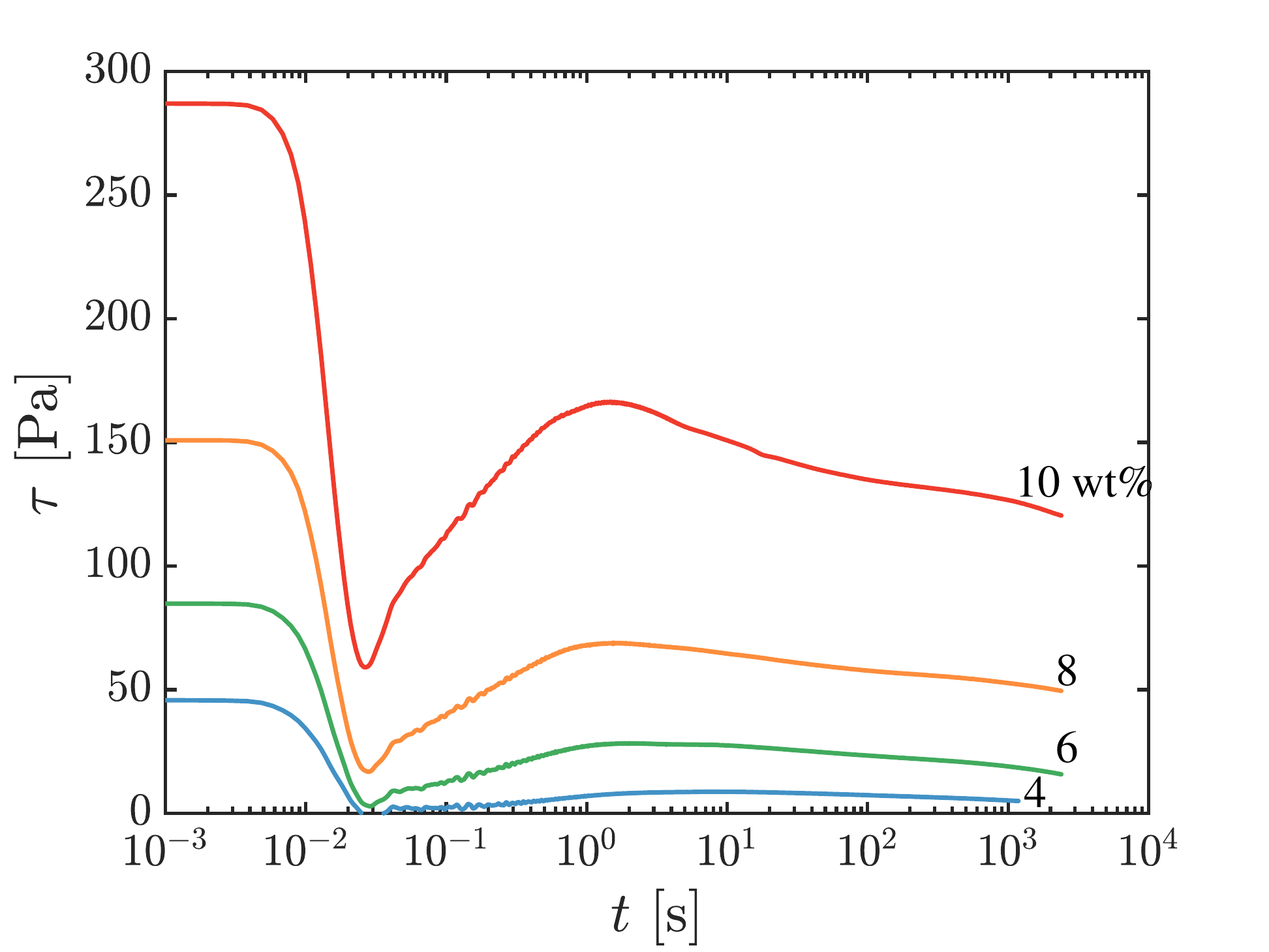}
	\end{minipage}
	\begin{minipage}[!h]{0.3\textwidth}
		\centering
		\textbf{(b)}\par
		\includegraphics[scale=0.27,trim={0 0 0 0},clip]{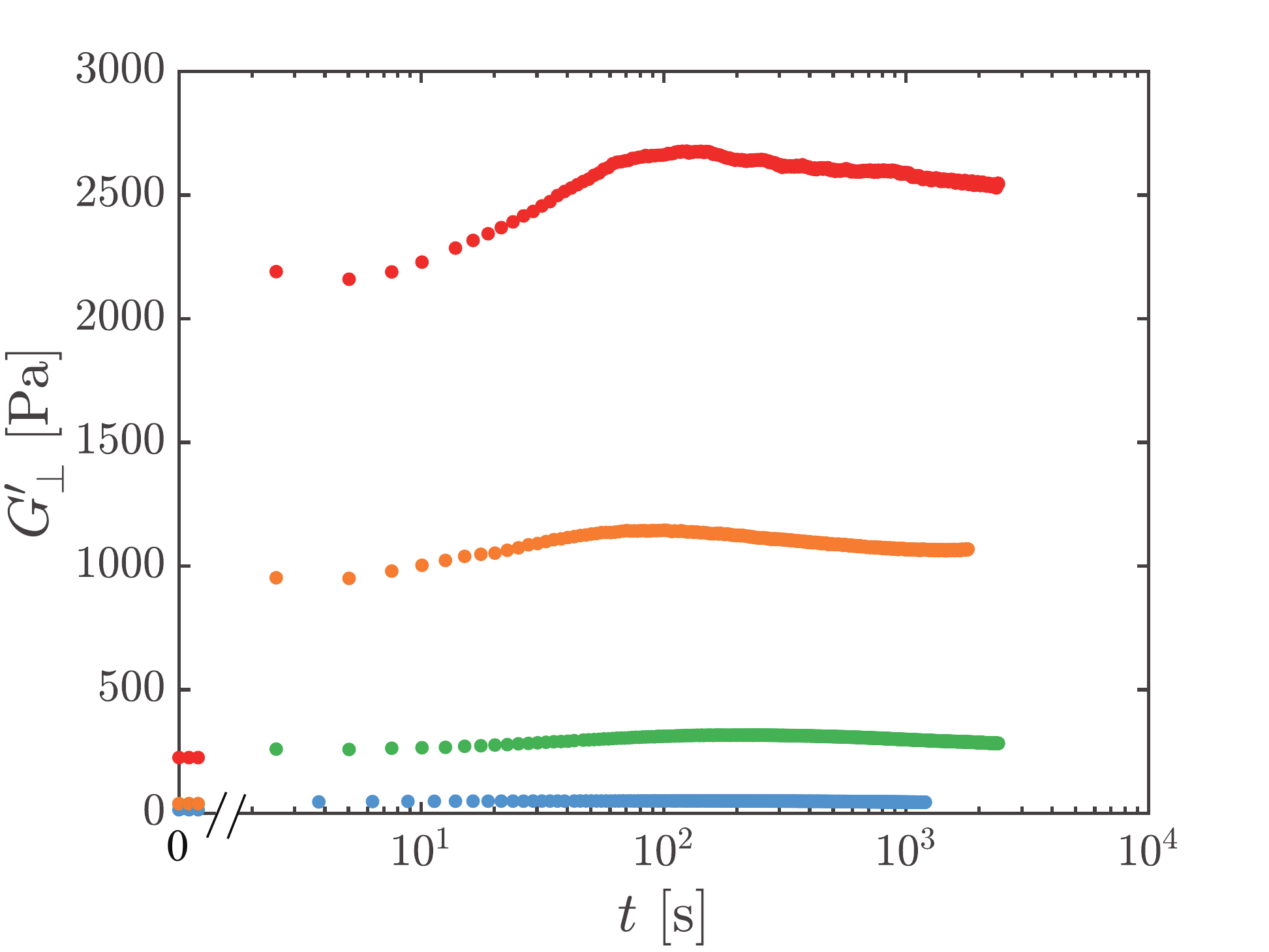}
	\end{minipage}
	\begin{minipage}[!h]{0.3\textwidth}
		\centering
		\textbf{(c)}\par
		\includegraphics[scale=0.27,trim={0 0 0 0},clip]{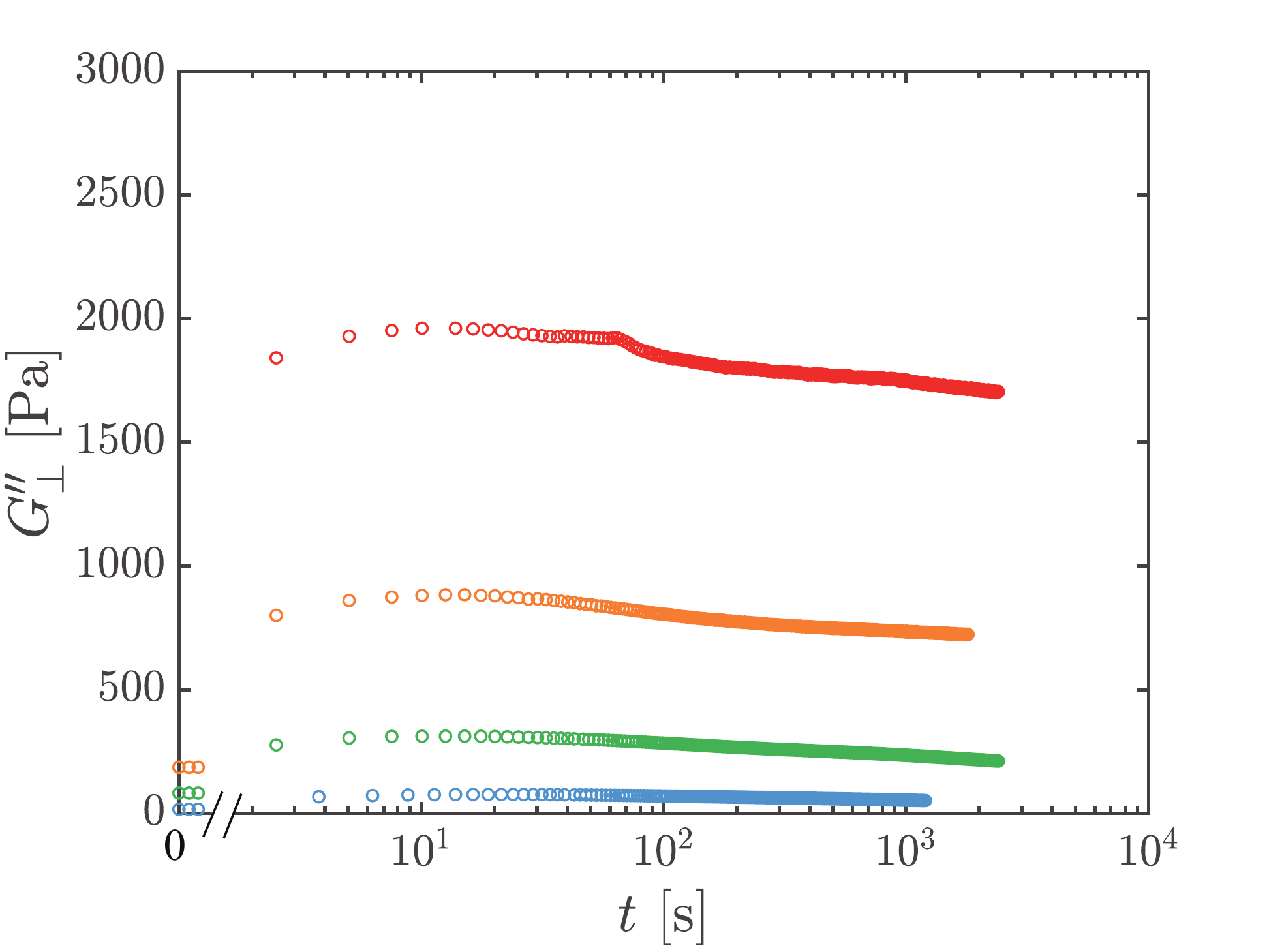}
	\end{minipage}
	\caption{The combined short-time thixotropic recovery and long-time anti-thixotropic decay in shear stress and orthogonal moduli is ubiquitous in the four different concentrations of CB suspensions tested, which are 4, 6, 8, and 10~wt$\%$. The (a) shear stress, (b) orthogonal storage modulus, and (c) the orthogonal loss modulus for the four concentrations of CB suspensions are plotted as a function of time during a step-down test, where the shear rate is decreased from 100 to 1~$\rm s^{-1}$.}
	\label{Fig.VulcanCB_concentration_dependence}
\end{figure*}

The combined short-time thixotropic recovery and long-time anti-thixotropic decay in shear stress and orthogonal moduli are ubiquitous across all four concentrations of CB suspensions tested (4, 6, 8, and 10~wt$\%$). Fig.~\ref{Fig.VulcanCB_concentration_dependence} shows the shear stresses, orthogonal storage moduli, and orthogonal loss moduli in response to a step down shear from 100 to 1~$\rm s^{-1}$ for CB suspensions of the four concentrations. The larger the CB concentration, the higher the shear stress, as well as the orthogonal moduli, as expected. Both the thixotropic recovery and anti-thixotropic decay are more dramatic for higher concentration suspensions, showing larger stress and storage moduli recovered. For the lowest concentration, 4~wt$\%$, the thixotropy and anti-thixotropy in shear stress and orthogonal moduli are more difficult to observe. 

There is no obvious difference among timescales for different concentrations. However, for all concentrations, similar to the 6~wt$\%$ suspension, as shown in Sec.~\ref{sec:timescale}, the difference in timescales of the shear response and the orthogonal response is distinct, with thixotropic times for orthogonal moduli orders of magnitude longer than those for shear stress. The anisotropy in the thixotropic timescale is consistent with the existence of vorticity aligned log-rolling structures in CB suspensions over a wide range of concentrations, as discussed in Sec.~\ref{sec:timescale}. 

\subsection{\label{sec:hysteresis}Additional evidence for anti-thixotropy from hysteresis loops}
In this section, we show additional evidence of anti-thixotropy of CB suspensions using hysteresis tests. There are multiple ways of conducting a hysteresis test \cite{Divoux2013, Greener1986, Bird1968, Marsh1968}. In this study, the hysteresis flow was applied on CB suspensions by starting at high shear rate and decreasing it from 300 to 0.03~$\rm s^{-1}$ through $n=20$ successive logarithmically spaced steps per decade of $\dot \gamma$ with duration $\delta t$ at each step, and then increasing to 300~$\rm s^{-1}$ again following the same steps and the same duration $\delta t$ per step. In this way, the total number of shear rates in a decreasing ramping was $N = 81$ steps. Different duration per step, $\delta t$, were used \cite{Divoux2013}, which were $\delta t=1$, 5, 20, 50, and 150~s. 

\begin{figure}
	\includegraphics[scale=.4]{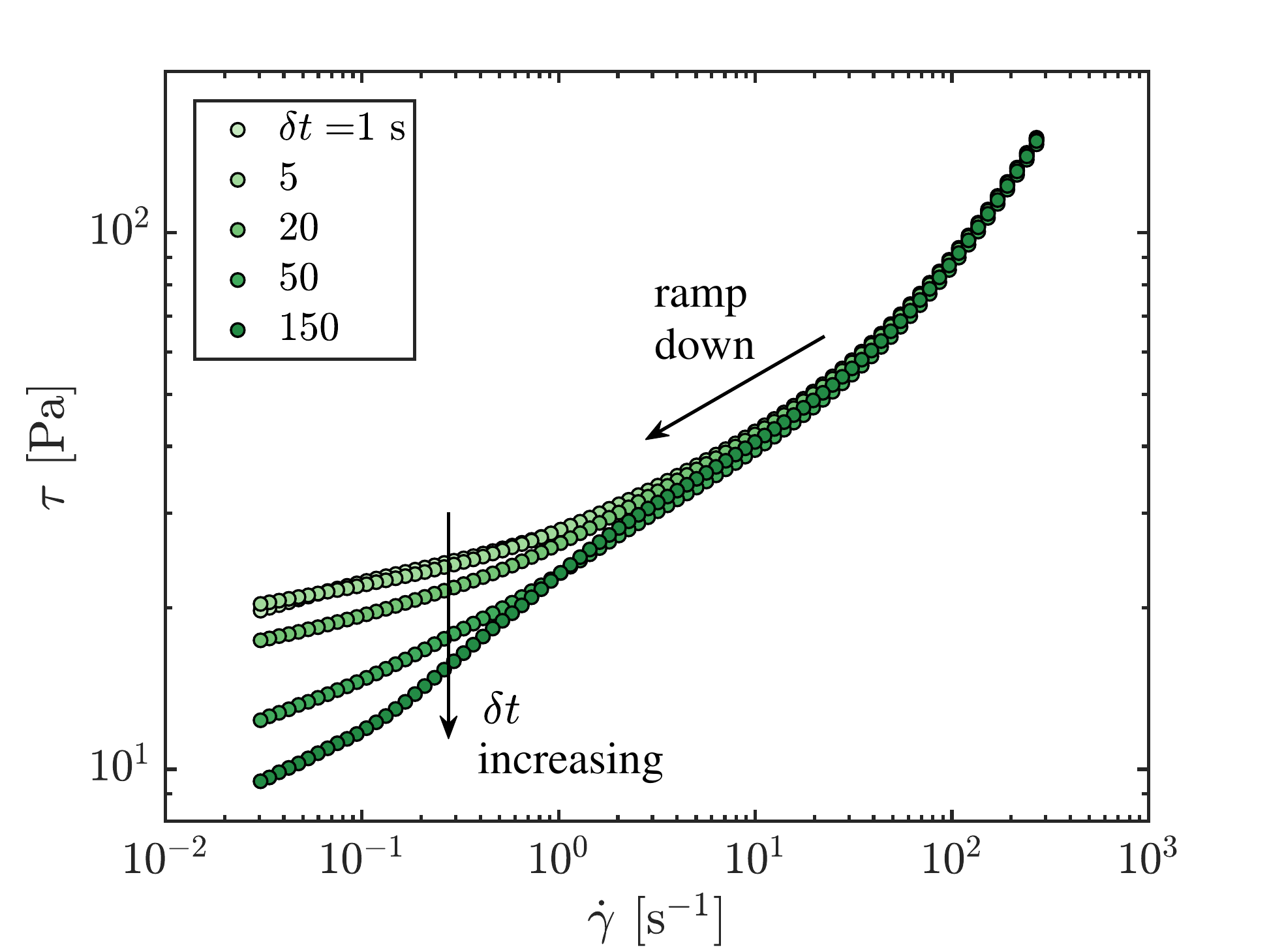}	
	\caption{Shear stress versus shear rate in first-half of hysteresis loop (ramping high-to-low) shift down with duration, $\delta t$, measured in a 6~wt$\%$ CB suspension, where shear rate is decreased from 300 to 0.03~$\rm s^{-1}$ with different duration. Each ramp is performed with $n$ = 20 points per decade, and a different duration per point $\delta t$ is used for each experiment. The arrow indicates the direction of ramping (from high to low).}
	\label{Fig.VulcanCB_6pctwt_hysteresis_stress_rate}
\end{figure}

Fig.~\ref{Fig.VulcanCB_6pctwt_hysteresis_stress_rate} shows the first half of the hysteresis loops for 6~wt$\%$ CB suspension, that is, when the shear rate ramps down from 300 to 0.03~$\rm s^{-1}$. The hysteresis curve with $\delta t=1~\rm s$ is shown in light green, and the curve with longest duration $\delta t=150~\rm s$ is shown in dark green. For quick ramping, where the duration is $\delta t=1~\rm s$, the total ramping time is 80~s. In this case, the CB structure is not given enough time to restructure and the suspension behaves like a yield stress fluid with a Herschel-Bulkley shaped flow curve. With the increase in duration, the CB structure has more time to rearrange at each step, and the stress decreases more at each shear rate. For the longest duration, $\delta t=150~\rm s$, the total ramping time is 12,000~s, and the stress is the lowest, with minimal apparent dynamic yield stress.  
 
\begin{figure}
	\includegraphics[scale=.4]{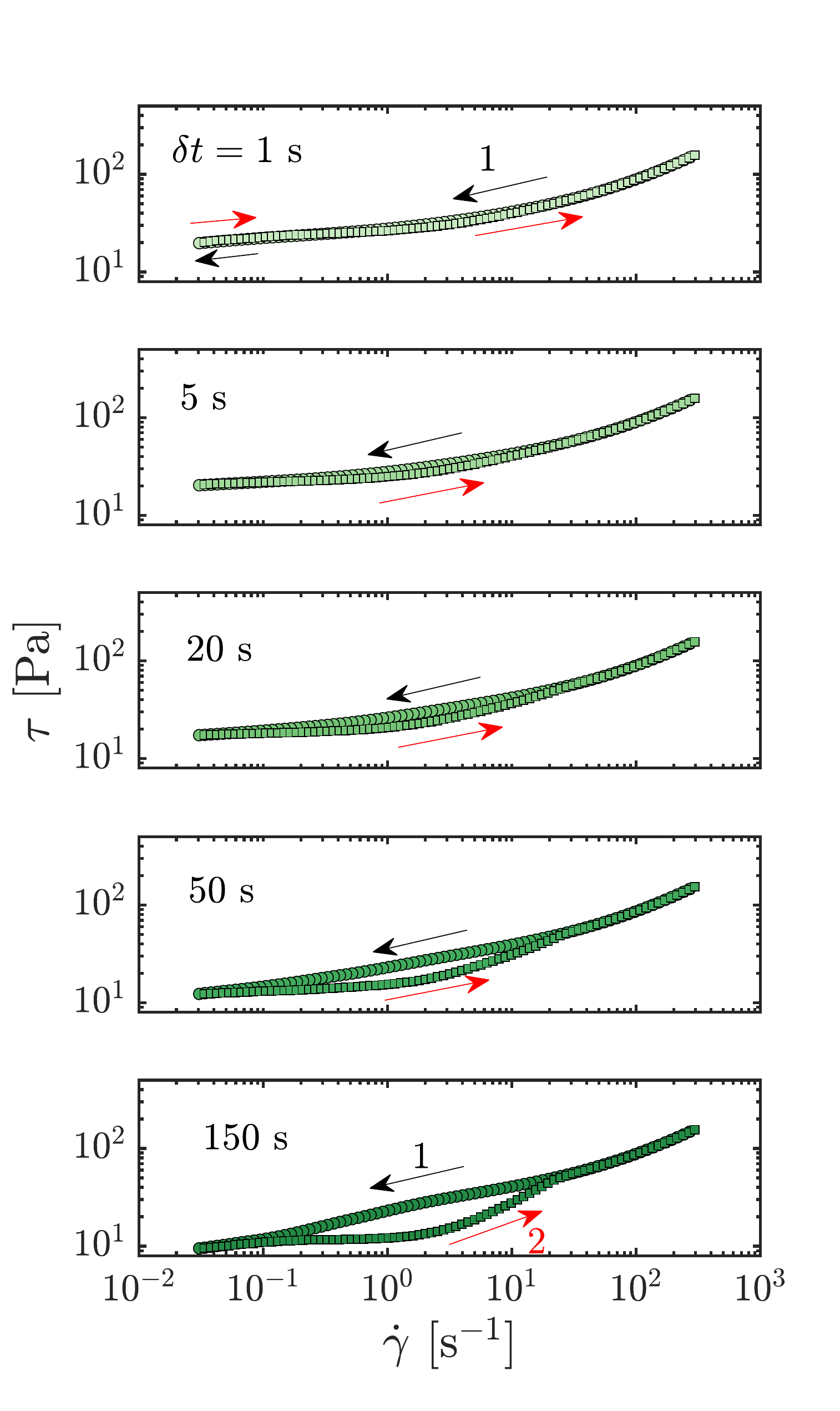}	
	\caption{Hysteresis tests indicate anti-thixotropic dynamics with counter-clockwise loops. Hysteresis loops measured in a 6~wt$\%$ CB suspension, by decreasing shear rate from 300 to 0.03~$\rm s^{-1}$ and then increasing it with the same step. Each ramp is performed with $n$ = 20 points per decade, and a different duration per point $\delta t$ is used for each experiment.}
	\label{Fig.VulcanCB_6pctwt_hysteresis}
\end{figure}

 Another feature of the hysteresis tests of CB suspensions is the direction of the hysteresis loops, which consist of both the downward and upward ramping. Fig.~\ref{Fig.VulcanCB_6pctwt_hysteresis} shows the hysteresis loops with different duration, from top to bottom, the duration per step increasing from 1 s to 150~s, and the total time of the hysteresis loop increasing proportionally. The data plotted as circles represents the downward ramping, labeled with the black arrow and the number "1" shown in black, while the squares, the red arrow, and the number "2" represent the upward ramping. From the plot we can see that for quick ramping, where $\delta t=1~\rm s$, the up and down ramps overlap and the area of the loop is vanishingly small. This is because during the hysteresis test (total time of 160~s, including both up and down ramping), the CB structure is not given enough time to restructure and therefore, the rheological properties of the suspension remain the same in down and up ramping. With the increase in $\delta t$, the area of the hysteresis loop increases. The increase in the area of hysteresis loops with duration is consistent with previous studies by Divoux and coworkers \cite{Divoux2013, Jamali2019}, where they found that for thixotropic materials, the area first increases then decreases with the duration, and the duration for the maximum area provides a timescale that can be compared between different materials. However, the previous studies on hysteresis loops fail to point out the direction of the loops, which is an important feature. The direction of the hysteresis loop of CB suspensions, if there is any, is counter-clockwise: the downward ramping stress is higher than the upward ramping one.

In our recent studies, we show that there are two distinguishing features in hysteresis loops that can fully differentiate between thixotropy, anti-thixotropy, and viscoelasticity \cite{Wang2021}. The direction of the hysteresis loop for thixotropic materials is clockwise, while for anti-thixotropic and viscoelastic materials, they are counter-clockwise. The second feature is that at quick ramping (small duration $\delta t$), where the down- and upward ramping flow curves overlap, the stress is a constant for a viscoelastic fluid that cannot relax, while the flow curve suggests a yield stress fluid for thixotropic and anti-thixotropic materials. For the CB suspensions tested here, the direction of hysteresis loops is counter-clockwise, while the stress rate curve shows a Herschel-Bulkley shape at quick ramping ($\delta t=1~\rm s$). This adds additional evidence that the CB suspension is anti-thixotropic. It is worth noticing that for the shortest duration ($\delta t = 1\rm ~s$), the upward ramping stress is slightly higher than the downward ramping stress, showing a clockwise loop at the low shear rate regime. This is consistent with the short-time thixotropic dynamics, as shown in the stress recovery in Fig.~\ref{Fig.VulcanCB_6pctwt_stepshear_stress}.

\section{\label{sec:discussions}DISCUSSION}

\subsection{\label{sec:structure}Microstructural model to rationalize the anisotropic thixotropy and anti-thixotropy}

\begin{figure}
	\includegraphics[scale=.5]{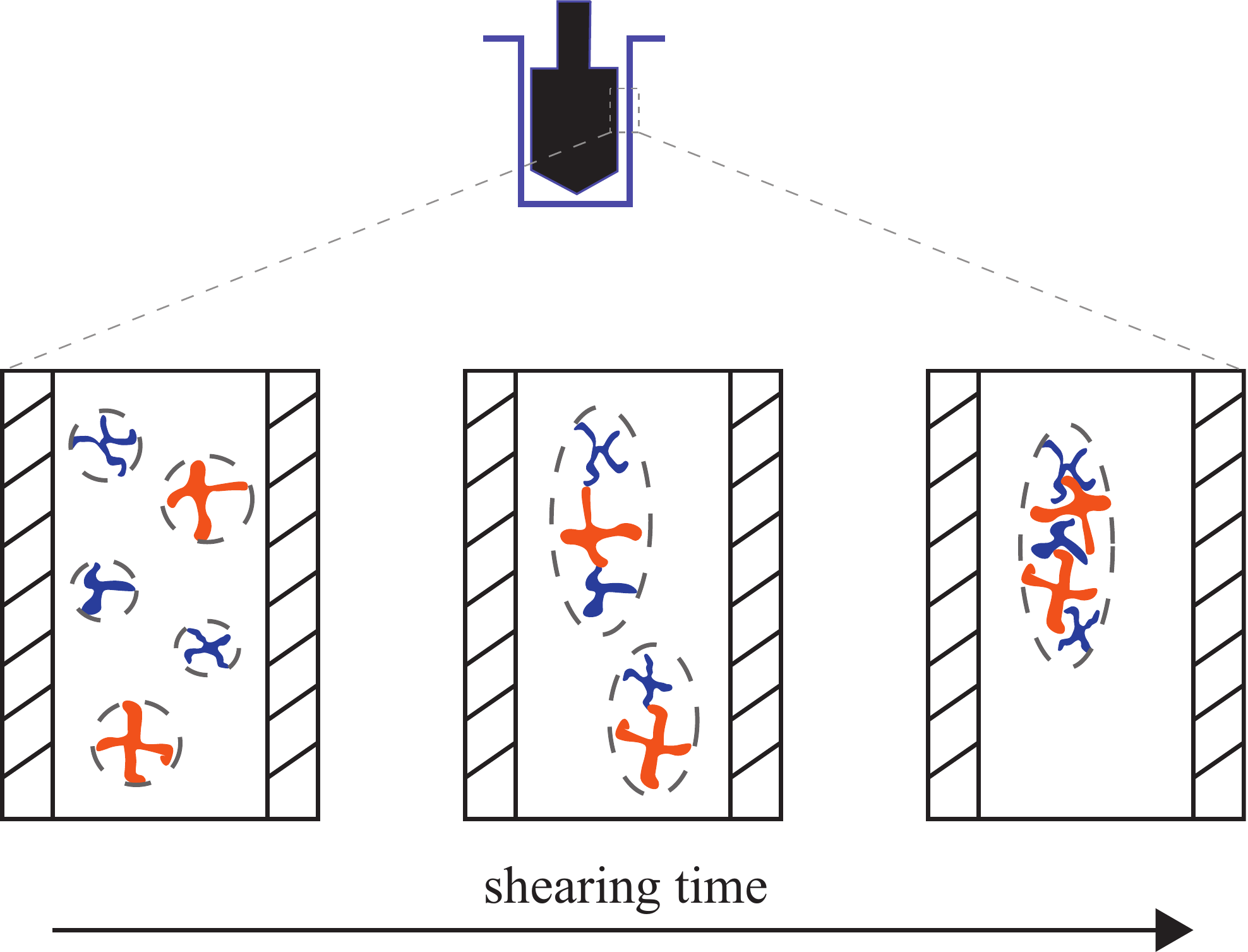}	
	\caption{Possible mechanical schematics of the CB structure under step-down shear rate flow. The open, large, more structured and loosely-connected agglomerates (middle panel) are formed during the short-time thixotropic build-up, which causes shear stress and orthogonal moduli to increase; while the shear-induced anti-thixotropic densification (right panel) results in stress decay at long times. The structure is anisotropic, and being longer in the vorticity direction, and rotates around the vorticity axis under shear. A concentric cylinder geometry is shown in the schematic for simplicity, whereas a double-wall concentric cylinder is used in experiments.}
	\label{Fig.CB_mechanical_schematics}
\end{figure}

We hypothesize a microstructural schematic to rationalize the anisotropic thixotropic and anti-thixotropic dynamics of CB suspensions under step-down shear rate flow, as shown in Fig.~\ref{Fig.CB_mechanical_schematics}. Our extensive rheological test results suggest a two-timescales structure rearrangement process, which in short timescales increases the viscosity and moduli of the suspension, and in long timescales decreases them. Based on this, we conjecture that at short times, relatively open, large, and loosely-connected agglomerates form under low final shear rates due to attractive van der Waals forces between CB aggregates, resulting in an overall more structured state and a higher effective hydrodynamic volume fraction. This increases the viscosity of the suspension and the resulting shear stress. At such low shear rate, log-rolling structures begin to form, which has been observed and reported for carbon black suspensions before \cite{Varga2019}. Therefore, we hypothesize that the structured agglomerates are asymmetric in shape, aligning in the vorticity direction and rotating around the vorticity axis. The anisotropy in structure results in the difference in the timescales between shear stresses and orthogonal moduli, and the difference in the magnitude of the moduli in the parallel and orthogonal directions. At longer times, the observed anti-thixotropic dynamics is due to these open agglomerates self-organizing and inter-penetrating, condensing the flocs into a denser, strongly-connected, and closed structure in both shear and orthogonal directions. This decreases the hydrodynamic volume, and therefore decreases the shear stress and moduli over long time scale. Such densification is shear-induced, as evidenced by the lack of moduli evolution in the cessation flow in Fig.~\ref{Fig.VulcanCB_6pctwt_stepshear_coplot_100_0}, and this can explain the long-time anti-thixotropic decay. 

At high shear rates, however, the anti-thixotropic decay is negligible in both the shear stress and the orthogonal moduli, as shown in Fig.~\ref{Fig.VulcanCB_6pctwt_stepshear_stress} and Fig.~\ref{Fig.VulcanCB_6pctwt_OSP_stepshear}. This is because the short-term thixotropic structure build-up is small since the change in the state of structure is small before and after stepping. The large, open, and loosely-connected anisotropic agglomerates are more dense and have less void fraction compared with those formed under lower final shear rates. Therefore, the long-time anti-thixotropic densification is not significant, resulting in negligible anti-thixotropic decay over long time, and the stresses and orthogonal moduli reach steady states more quickly. The dynamics at large shear rates is consistent with the microstructural model shown in Fig.~\ref{Fig.CB_mechanical_schematics}.

\subsection{\label{sec:anisotropy}The growth of anisotropy under low shear}

\begin{figure*}[h!]
	\begin{minipage}[!h]{0.49\textwidth}
		\centering
		\textbf{(a)}\par
		\includegraphics[scale=1.1,trim={0 0 0 0},clip]{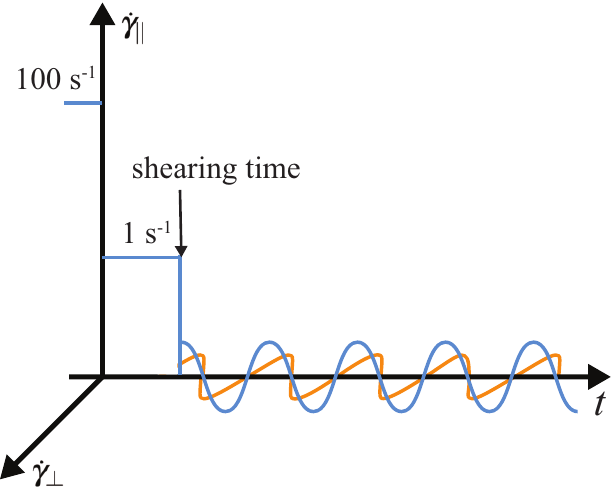}
	\end{minipage}
	\begin{minipage}[!h]{0.49\textwidth}
		\centering
		\textbf{(b)}\par
		\includegraphics[scale=0.55,trim={0 0 0 0},clip]{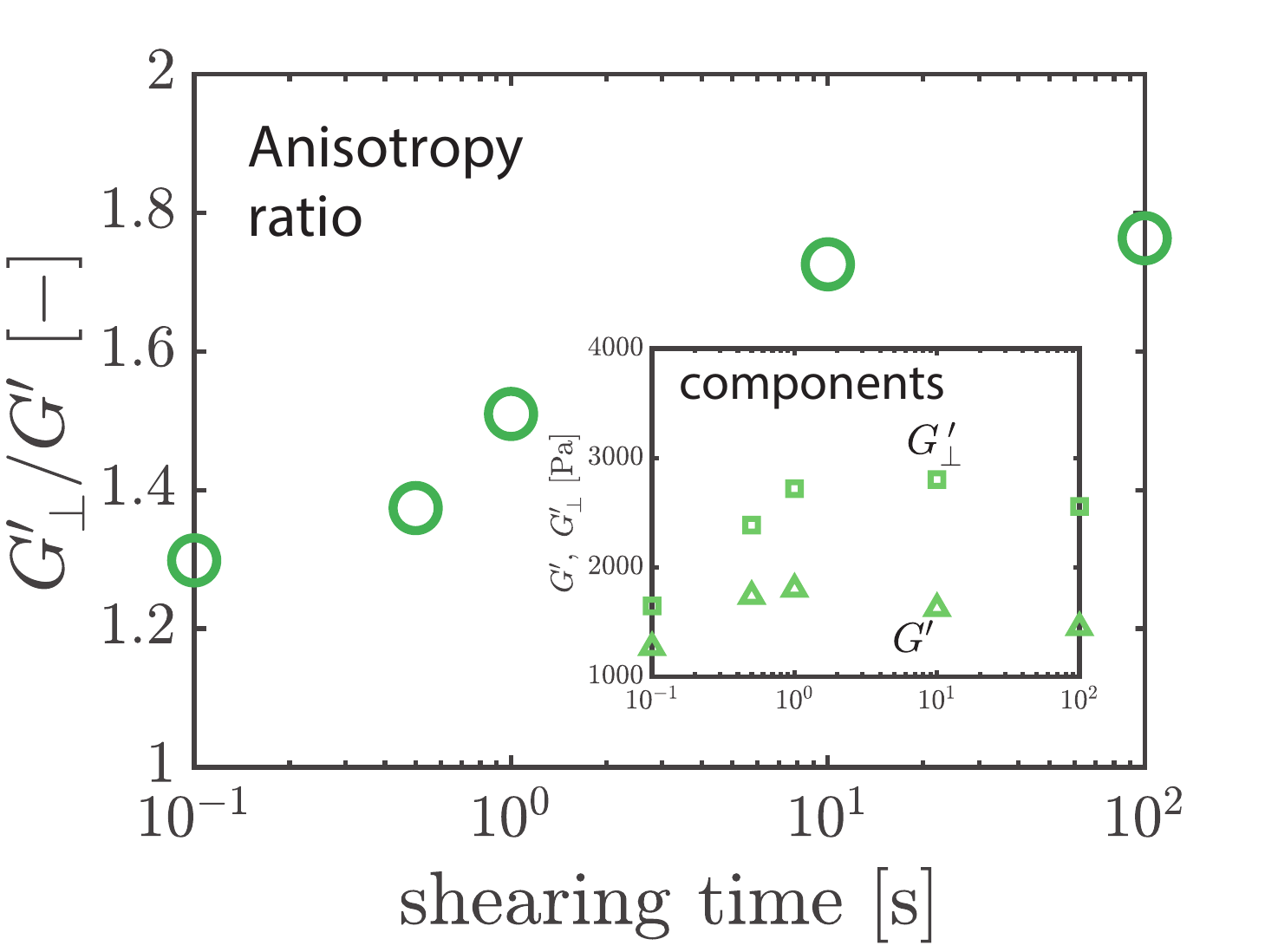}
	\end{minipage}
	\caption{The degree of anisotropy increases with the shearing time; the orthogonal and rotational elastic moduli both show the thixotropic recovery and anti-thixotropic decay during the step-down shear test (from 100 to 1~$\rm s^{-1}$), but their ratio keeps increasing.}
	\label{Fig.anisotropy}
\end{figure*}

The mechanical schematics in Fig.~\ref{Fig.CB_mechanical_schematics} show how anisotropic structure grows qualitatively; in this section, we will show the growth of anisotropic structure more quantitatively using 2D-SAOS. Fig.~\ref{Fig.anisotropy} shows the applied shear input on the CB suspension, where the sample is sheared at $1~\rm s^{-1}$ for a given shearing time, after initially reaching steady state at 100~$\rm s^{-1}$. Then the flow is ceased and 2D-SAOS at 0.5$\%$ peak strain amplitude and frequency 100~rad/s is imposed to simultaneously measure $G^{\prime}$ and $G^\prime_\perp$ as a function of the shearing time. To quantify the degree of anisotropy, we define an anisotropy parameter, $G^\prime_\perp / G^{\prime}$, the ratio of the orthogonal and parallel storage moduli. The storage moduli and their ratio are plotted as functions of the shearing time, as shown in Fig.~\ref{Fig.anisotropy}(b). Both shear and orthogonal moduli show thixotropic recovery and anti-thixotropic decay, with moduli first increasing then decreasing with the shearing time. This is consistent with what we observed before in Fig.~\ref{Fig.VulcanCB_6pctwt_stepshear_stress} and Fig.~\ref{Fig.VulcanCB_6pctwt_OSP_stepshear}. However, the ratio of the orthogonal and shear moduli, $G^\prime_\perp / G^{\prime}$, keeps rising with the shearing time, increasing by 33$\%$ in 10~s and reaching a constant ratio of 1.8. The change in both the elastic moduli and their ratio suggests that along with the thixotropic structure build-up and anti-thixotropic decay, the anisotropic structure in CB suspensions under low shear keeps evolving and increasing over these timescales. 

\section{\label{sec:conclusions}CONCLUSIONS}
In summary, this study resolved the long-existing ambiguity of transient dynamics of CB suspensions by using OSP during step-down shear flows, and this is the first time OSP has been used to study the anti-thixotropic shear-induced structuring of CB suspensions. From step-down flow tests on CB in mineral oil suspensions at different concentrations from 4 to 10~wt$\%$, a two-region stress response is identified. In the high shear rate region, stress shows a negligible thixotropic recovery and reaches steady state quickly. While in the low shear rate region, it shows a short-time thixotropic recovery and a long-time decay. The combined recovery-then-decay was verified to be a reversible, reproducible, and intrinsic material response, and the long-time decay was attributed to anti-thixotropy, rather than viscoelasticity with the help of OSP. This demonstrates that the response involves shear-induced structuring that affects both viscous and elastic properties. We further showed a mechanical anisotropy in the CB suspensions under shear, where we found that the decay in the orthogonal direction happens much later than that in the rotational direction, and the orthogonal moduli are two times larger than the moduli in the rotational direction. Such mechanical anisotropy suggests the existence of an anisotropic log-rolling structure aligned in the vorticity direction during the stepping-down in shear flow. A mechanical structure schematic was proposed, considering qualitatively thixotropic structure build-up and anti-thixotropic densification and anisotropic log-rolling structure. The growth of anisotropy was verified by the moduli measured in both directions using OSP 2D-SAOS. 

It should be noted that a steady state was never observed for the transient shear stress, or the orthogonal storage and loss moduli of all suspensions even after 20 minutes of shearing, once they begin to show anti-thixotropic decay. The orthogonal moduli measured using OSP clarify that such decay is anti-thixotropic and can possibly be attributed to shear-induced anti-thixotropic structure densification at long times. 

Importantly, the results presented here suggest a combined thixotropic and anti-thixotropic dynamics that happens at different timescales, which has not been acknowledged before. This is the first time the thixotropy, anti-thixotropy, and viscoelasticity of CB suspensions are clearly differentiated, and using OSP combined with step shear rate tests provides a protocol to resolve the ambiguity between anti-thixotropy and viscoelasticity. Our observation for these CB suspensions is outside the standard paradigm of thixotropic structure-parameter constitutive models, and this means that the existing kinetic thixotropic models with only one structure parameter to represent the structure build-up state are not enough to describe the structure change of CB under even simple shear flow. The microstructural model of CB we proposed can rationalize the mechanical responses of CB suspensions, which can be useful in designing flow-related devices, such as semi-solid flow batteries, where CB is used as a conductive additive.

\bibliographystyle{unsrt}
\bibliography{CB}

\end{document}